\documentclass[amsmath,amssymb,reqno]{amsart}

\numberwithin{equation}{section}

\usepackage{bbm,amsthm,amsmath,amssymb,times}

\newtheorem{theorem}{Theorem}[section]
\newtheorem{proposition}[theorem]{Proposition}
\newtheorem{lemma}[theorem]{Lemma}
\newtheorem{corollary}[theorem]{Corollary}
\theoremstyle{definition} 
 
\newtheorem{example}[theorem]{Example}
\newtheorem{remark}[theorem]{Remark}
\newtheorem{remarks}[theorem]{Remarks}
\newtheorem{definition}[theorem]{Definition}

\makeatletter

\def\ps@firstpage{\ps@plain
  \def\@oddhead{
    \begin{minipage}{\textwidth}
      \normalfont\scriptsize To appear in: {\it Annales Henri Poincar\'e} \hfill 
    \end{minipage}}}%

\def\@tocpagenum#1{\hss{\mdseries #1}}
\def\@tocwrite#1{\@xp\@tocwriteb\csname toc#1\endcsname{#1}}
\def\@tocwriteb#1#2#3{%
  \begingroup
    \def\@tocline##1##2##3##4##5##6{%
      \ifnum##1>\c@tocdepth
      \else \sbox\z@{##5\let\indentlabel\@tochangmeasure##6}\fi}%
    \csname l@#2\endcsname{#1{\csname#2name\endcsname}{\@secnumber}{}}%
  \endgroup
  \addcontentsline{toc}{#2}%
    {\protect#1{\csname#2name\endcsname}{\@secnumber}{#3}}}
\def\l@section{\@tocline{1}{0pt}{0.5pc}{}{}}
\renewcommand{\tocsection}[3]{%
  \indentlabel{\@ifnotempty{#2}{\ignorespaces#1 #2.\quad}}#3}
\def\l@subsection{\@tocline{2}{0pt}{2.5pc}{5pc}{}}

\def\l@subsubsection{\@tocline{3}{0pt}{1pc}{7pc}{}}

\def\l@part{\@tocline{-1}{12pt plus2pt}{0pt}{}{\bfseries}}

\def\l@chapter{\@tocline{0}{8pt plus1pt}{0pt}{}{}}

\makeatother
\RequirePackage{multicol} 

\makeatletter

\renewcommand\indexname{Citation Index}

\def\citationindex{
                \@input@{\jobname.ind}}%

\makeindex

\def\@citex[#1]#2{%
  \let\@citea\@empty
  \@cite{\@for\@citeb:=#2\do%
         {\@citea\def\@citea{,\penalty\@m\ }%
          \edef\@citeb{\expandafter\@firstofone\@citeb}%
          \if@filesw\immediate\write\@auxout{\string\citation{\@citeb}}\fi%
          \@ifundefined{b@\@citeb}%
          {\mbox{\reset@font\bfseries ?}%
             \G@refundefinedtrue\@latex@warning
             {Citation `\@citeb' on page \thepage \space undefined}}%
          {\hbox{\csname b@\@citeb\endcsname}%
            \def\speicher{\csname b@\@citeb\endcsname}%
            \makeatother%
                \index{\speicher @[\speicher]}%
            \makeatletter%
          }
         }
        }{#1}
}%

\def\@idxitem#1,{\par\noindent#1\quad\hfill\raggedleft\hangindent 5\p@}

\newcounter{numcount}
\newcommand{\labelnummer}{\mbox{(\roman{numcount})}}%

    {\let\curlabelspeicher\@currentlabel%
     \begin{list}{\labelnummer}{\usecounter{numcount}%
                  \topsep1ex\partopsep2ex\parsep0pt\itemsep1.5ex\@plus1\p@%
                  \labelwidth3em\itemindent0em\labelsep1em%
                  \leftmargin3em}%
     \let\saveitem\item%
     \def\item{\saveitem%
               \def\@currentlabel{\curlabelspeicher\labelnummer}%
               \let\label\bemlabel}}%
   {\end{list}}%

\newenvironment{indentnummer*}%
    {\begin{list}{\labelnummer}{\usecounter{numcount}%
                  \topsep1ex\partopsep2ex\parsep0pt\itemsep1ex
                  \labelwidth2.5em\itemindent0em\labelsep1em%
                  \leftmargin3.5em
                  }}%
   {\end{list}}%

\newenvironment{nummer}%
    {\let\curlabelspeicher\@currentlabel%
     \begin{list}{\labelnummer}{\usecounter{numcount}\leftmargin0em%
                  \topsep1ex\partopsep2ex\parsep0pt\itemsep0.5ex
                  \labelwidth2.5em\itemindent3.5em\labelsep1em}%
     \let\saveitem\item%
     \def\item{\saveitem%
               \def\@currentlabel{\curlabelspeicher\labelnummer}%
               \let\label\bemlabel}}%
   {\end{list}}%

\def\itemref#1{\expandafter\@setref\csname r@#1item\endcsname\@firstoftwo{#1}}%

\def\bemlabel#1{\@bsphack%
  \protected@write\@auxout{}%
         {\string\newlabel{#1}{{\@currentlabel}{\thepage}}}%
  \ifmmode\else%
  \protected@write\@auxout{}%
         {\string\newlabel{#1item}{{\labelnummer}{\thepage}}}%
  \fi%
  \@esphack}%

\begin{document}
\bibliographystyle{plain}

\title[Energetic and dynamic properties]{Energetic and dynamic properties of a \\ quantum particle  
in a spatially random magnetic field \\ with constant correlations along one direction}

\author{Hajo Leschke} 
\address{%
Institut f\"ur Theoretische Physik, Universit\"at Erlangen-N\"urnberg, 
Staudtstr.\ 7, 91058 Erlangen, Germany
}%
\author{Simone Warzel}%
\address{%
Jadwin Hall, Princeton University, NJ 08544, USA. {\rm On leave from:} 
Institut f\"ur Theoreti\-sche Physik, Universit\"at Erlangen-N\"urnberg,
Staudtstr.\ 7, 91058 Erlangen, Germany
}%
\author{Alexandra Weichlein}
\address{%
Institut f\"ur Theoretische Physik, Universit\"at Erlangen-N\"urnberg, 
Staudtstr.\ 7, 91058 Erlangen, Germany 
}%

\date{July, 15, 2005}

\begin{abstract}
 We consider an electrically charged particle on the Euclidean plane subjected to a perpendicular magnetic field which depends only on one
 of the two Cartesian co-ordinates.
 For such a ``unidirectionally constant'' magnetic field (UMF), which otherwise may be random or not, we prove certain
 spectral and transport properties associated with the corresponding one-particle Schr\"odinger operator (without scalar potential) 
 by analysing 
 its ``energy-band structure''. 
 In particular, for an ergodic random UMF we provide conditions 
 which ensure that the operator's entire spectrum is almost surely absolutely continuous. 
 This implies that, along the direction in which the random UMF is constant,
 the quantum-mechanical motion is almost surely ballistic, while in the perpendicular direction in the plane one has 
 dynamical localisation.
 The conditions are verified, for example, for Gaussian and Poissonian random UMF's with non-zero mean-values. 
 These results may be viewed as ``random analogues'' of results first 
 obtained by A.\ Iwatsuka [Publ.\ RIMS, Kyoto Univ.\ {\bf 21} (1985) 385]
 and (non-rigorously) by J.\ E.\ M\"uller [Phys.\ Rev.\ Lett.\ {\bf 68} (1992) 385].\\
 
 \begin{center}
   \begin{minipage}{6cm}
     In memoriam\\
     Heinz BAUER (31 January 1928 -- 15 August 2002)\\
     former Professor of Mathematics at the University of Erlangen-N\"urnberg
   \end{minipage}
 \end{center}
\end{abstract}


\maketitle
\tableofcontents

\section{Introduction}\label{Intro}
The quantum-dynamical behaviour of electrically charged particles in a spatially
\emph{random magnetic field} (RMF) has become a topic of growing interest
over the last decade.  Most theoretical investigations of corresponding
one-particle models take their motivation from the physics of (quasi-) 
two-dimensional systems.  For example, in connection with the fractional quantum Hall
effect, transport properties of interacting electrons 
on the (infinitely-extended) Euclidean plane $ \mathbb{R}^2 $ subjected to an external random scalar potential and a
perpendicular, strong homogeneous magnetic field are often
described by (non-interacting) effective, so-called   
composite fermions in a RMF, which is
homogeneous on average.  Near half filling of the lowest Landau level,
the values of this (fictitious) RMF fluctuate at each point 
$x = (x_1 , x_2 ) \in \mathbb{R}^2 $ about a mean-value near zero
\cite{Hei98,Wol00,MuSh03}. Moreover, experimental realisations of gases of 
non-interacting fermions in (actual) RMF's by quasi-two-dimensional semiconductor
heterostructures with certain randomly built-in magnets have been
reported \cite{GBGB94,Smi94,Man95,AAKI00,Rus00,Byk01,Rus03}.  
Last but not least, there is a fundamental
interest in the theory of one-particle models with RMF's in two
dimensions.  Just like in Anderson's problem \cite{And58} of a quantum
particle subjected to a random scalar potential (only), an important question is whether all
(generalised) energy eigenstates are spatially localised or whether
some of them are delocalised. Until recently, in the RMF-case the answer
to the question has remained controversial within perturbative,
quasi-classical, field-theoretical and numerical studies 
\cite{ArMi94,LeCh94,KaOh95,ShWe95,YaBa96,BaSc98,Eve99,Fur99,PoSc99,KaKr99,ShWe00,Yak00,TaEf01,Ngu02,EfKo03,KaOh03}.  
It is therefore desirable to establish exact localisation/delocalisation results
for the RMF-case as has been done for random scalar potentials \cite{CaLa90,PaFi92,Sto01} (see also \cite{LeMuWa03}).  
For the RMF-case (without a random scalar potential) 
we are aware of only one rigorous work \cite{KlNa02} devoted to the localisation/delocalisation problem.
Therein Klopp, Nakamura, Nakano and Nomura outline
a proof of the existence of localised states at low energies in a
certain model for a particle on the (unit-) square lattice $ \mathbb{Z}^2 $ instead of the two-dimensional continuum $ \mathbb{R}^2 $.

In the present paper we prove first exact localisation/delocalisation results for a simplified model for a particle
on the continuum $ \mathbb{R}^2 $.  
The simplification arises from the assumption that the
fluctuations of the RMF on $ \mathbb{R}^2 $ are anisotropically
long-ranged correlated in the sense that we consider the limiting case
of an infinite correlation length along one direction and take the
correlation length to be finite but strictly positive along the
perpendicular direction in the plane.  In other words, the RMF is
assumed to be independent of one of the two Cartesian coordinates, which
we choose to be the second one, $ x_2 $.  The remaining dependence of
the RMF-values on the first
coordinate $ x_1 $ we suppose to be governed by the realisations of an \emph{ergodic} real-valued 
random process with the real line $ \mathbbm{R} $ as its parameter set. 
For the precise description of such a \emph{random unidirectionally constant magnetic field} (RUMF) 
see Definition~\ref{assr} below. 
To our knowledge, the first rigorous work explicitly dealing with a model involving a random UMF (with zero mean-value) 
is one of Ueki \cite{Uek00}.

Models for a single particle on the plane $\mathbbm{R}^2$ subjected to a non-random \emph{unidirectionally constant magnetic field} (UMF) have been the object of various studies 
in the mathematics \cite{Iwa85,CyFr87,MaPu97,LeRu02} and physics \cite{Mul92,NoBe92,Kra95,RePe00,Sim01,Law02} literature. 
These models illustrate unadulteratedly that inhomogeneous magnetic fields have a tendency to delocalise charged particles 
along the direction perpendicular to the magnetic-field gradient. According to classical mechanics 
a particle with non-zero kinetic energy wanders off to infinity along snake or cycloid-like orbits 
winding around contours of constant magnetic field \cite{CyFr87,Mul92}. 
The quantum analogue of this unbounded motion should manifest itself in the exclusive appearence of 
absolutely continuous spectrum of the underlying
one-particle Schr\"odinger operator with a UMF (only), which is not globally constant.   
Although plausible from the (quasi{-)} classical picture, 
a mathematical proof of this conjecture is non-trivial and has been accomplished 
so far only for certain classes of UMF's \cite{Iwa85,MaPu97}.
From the same picture, the absolutely continuous spectrum should come with ballistic transport along the direction perpendicular to
the gradient of the UMF. 
Along the direction parallel to the gradient no propagation is expected, provided the UMF is non-zero on spatial average -- like in the case of a globally constant magnetic field. 

In the second section of the present paper we compile rigorous results on spectral and transport properties of one-particle Schr\"odinger operators with UMF's 
which are non-zero on spatial average. As far as transport is concerned, these results slightly extend the ones in \cite{MaPu97}.
In the third and main section we formulate conditions on the RUMF which imply 
that the spectrum of the corresponding random Schr\"odinger operator is almost surely only absolutely continuous. 
By virtue of Section~\ref{sec:UMF} such a RUMF yields ballistic transport along one direction and dynamical 
localisation along the other almost surely. These results apply, for example, to Gaussian and Poissonian RUMF's 
with non-zero mean-values. 

Some of the results of the present paper have been announced in \cite{LeWa05}, where the key ideas are outlined only briefly.

\section{Schr{\"o}dinger operators with unidirectionally constant magnetic fields}\label{sec:UMF}
Throughout this section we are dealing with (non-random) unidirectionally constant magnetic fields in the sense of
\begin{definition}[UMF]\label{ass}
  A \emph{unidirectionally constant magnetic field} (UMF) 
  is given by a real-valued function $ b : \mathbbm{R} \to \mathbbm{R} $, $ x_1 \mapsto b(x_1) $, 
  which is locally Lebesgue-integrable, $ b \in {\rm L}^1_{\rm loc}(\mathbbm{R}) $, and
  whose anti-derivative
  \begin{equation} \label{eq:a}
    a: \mathbbm{R} \to \mathbbm{R}, \; x_1 \mapsto a(x_1):= \int_0^{x_1} \!\! \mathrm{d}y_1 \; b(y_1)
  \end{equation}
  behaves near infinity according to
  \begin{equation}\label{eq:anti}
    0 < \overline{b} :=  \liminf_{|x_1| \to \infty} \frac{|a(x_1)|}{|x_1|} \leq \infty \quad \mbox{and} 
    \quad 0 \leq \limsup_{|x_1| \to \infty} \frac{|a(x_1)|}{|x_1|^\alpha} < \infty 
    \quad \mbox{with some $ \alpha \geq 1$.}
  \end{equation} 
\end{definition}
Taking the function (\ref{eq:a}) as the second component of the vector potential $ \mathbbm{R}^2 \ni (x_1,x_2) $\hspace{0pt}$\mapsto ( 0, a(x_1) )$\hspace{0pt}$ \in \mathbbm{R}^2 $ 
in the asymmetric gauge, the Hamiltonian (or: Schr\"odinger operator) for a single spinless particle 
on the Euclidean plane $ \mathbb{R}^2 $ subjected to a UMF,
which depends (at most) on the first Cartesian co-ordinate $ x_1 $,
is informally given by the second-order differential operator
\begin{equation}\label{eq:H}
  H(b) := \frac{1}{2} \left[ P_1^2 + \left( P_2 - a(Q_1)\right)^2 \right].
\end{equation}
Here $ P_1:= -i \partial/\partial x_1 $, $ P_2:= -i \partial/\partial x_2 $ and $ Q_1 $(, $ Q_2 $) are the two
components of the canonical momentum, respectively, position operator on the Hilbert space $ {\rm L}^2(\mathbbm{R}^2) $ of 
complex-valued, Lebesgue square-integrable functions $\psi$ on $\mathbbm{R}^2$ with squared norm $\|\psi\|^2 $\hspace{0pt}$:=\int_{\mathbbm{R}^2} d^2x \, |\psi(x)|^2 < \infty $. 
The operators $Q_1$ and $Q_2$ act as multiplication by $ x_1 $, respectively, $ x_2 $.
Moreover, we use physical units in which Planck's constant (divided by $ 2 \pi $), the particle's mass and charge are all equal to $ 1 $.
The requirements in Definition~\ref{ass} guarantee that $ a $ is not only absolutely continuous and hence 
locally bounded, $ a \in {\rm L}^\infty_{\rm loc}(\mathbbm{R}) $, 
but also polynomially bounded near infinity. 
Therefore (\ref{eq:H}) is precisely defined as an essentially self-adjoint and non-negative operator on the Schwartz space 
$ \mathcal{S}(\mathbbm{R}^2) \subset {\rm L}^2(\mathbbm{R}^2) $ of complex-valued, arbitrarily often differentiable functions of rapid decrease near infinity (cf.~\cite[Thm.~1.15]{CyFr87}). In the context of quantum mechanics the operator (\ref{eq:H}) represents 
the total kinetic energy of the particle and generates its time evolution.

\subsection{Energy bands and related spectral properties}
Thanks to translation invariance along the $ x_2 $-direction the Hamiltonian (\ref{eq:H})
commutes with $ P_2 $ so that it may be fibred (or: decomposed) into the one-parameter family 
\begin{equation}\label{eq:redH}
  H^{(k)}(b) := \frac{1}{2} \left[ P_1^2 + ( k - a(Q_1) )^2 \right],\qquad k \in \mathbbm{R}
\end{equation}
of \emph{effective Hamiltonians} on the Hilbert space $ {\rm L}^2(\mathbbm{R}) $ for the one-dimensional 
motion along the $ x_1 $-direction, 
where each \emph{wave number} $ k \in \mathbbm{R} $ 
may be interpreted as a spectral value of $ P_2 $.  Definition~\ref{ass} implies that each 
$H^{(k)}(b) $ is essentially self-adjoint on $ \mathcal{S}(\mathbbm{R}) $.
The following proposition collects some well-known facts 
about the relations between $ H^{(k)}(b)$ and $ H(b) $ and their spectral properties. 
For its precise formulation we introduce the partial Fourier(-Plancherel) transformation $ \mathcal{F} $ given by 
\begin{equation}\label{eq:ParFour}
 \left( \mathcal{F}\psi\right)^{(k)}(x_1) := \frac{1}{\sqrt{2 \pi }} \int_{\mathbbm{R}} dx_2 \, e^{-i k x_2} \, \psi(x_1,x_2), \,\,\,\,\,\,  x_1 \in \mathbbm{R}
\end{equation}
for any $ \psi \in \mathcal{S}(\mathbbm{R}^2) $. It uniquely extends to a unitary operator 
$  \mathcal{F}: {\rm L}^2(\mathbbm{R}^2) \to  \int_{\mathbbm{R}}^\oplus \!\! dk \, {\rm L}^2(\mathbbm{R}) $ 
which maps onto the Hilbert space of $  {\rm L}^2(\mathbbm{R}) $-valued functions 
$  \mathcal{F}\psi : \mathbbm{R} \to {\rm L}^2(\mathbbm{R})$, $k \mapsto \left(\mathcal{F}\psi\right)^{(k)} $  
with Lebesgue square-integrable $  {\rm L}^2(\mathbbm{R}) $-norm, $ \int_\mathbbm{R}\! dk \, \big\| \left(\mathcal{F}\psi\right)^{(k)} \big\|^2 = \| \psi \|^2 < \infty $. 
\begin{proposition}[cf.~\cite{Iwa85,MaPu97}]\label{prop:spectral}
  Let $ b $ be a UMF. Then 
  \begin{nummer}
  \item
    the  family of operators $ \left\{ H^{(k)}(b) \right\}_{k \in \mathbbm{R}} $ is analytic of type A 
    (in the sense of \cite[Def.~on p.~16]{ReSi4}) in some complex neighbourhood of $ \mathbbm{R} $ . 
    For each fixed $ k \in \mathbbm{R} $ the spectrum of $ H^{(k)}(b) $ is only discrete and its spectral resolution reads 
    \begin{equation}\label{eq:sectral}
      H^{(k)}(b) = \sum_{n=0}^\infty  \varepsilon^{(k)}_n(b) \, E^{(k)}_n(b).
    \end{equation}
    The eigenvalues $ 0 < \varepsilon^{(k)}_0(b) <  \varepsilon^{(k)}_1(b) < \dots $ 
    are non-degenerate, strictly positive and analytic functions of $ k $ 
    in some complex neighbourhood of $ \mathbbm{R} $.
    By the non-degeneracy the corresponding orthogonal eigenprojections $ E^{(k)}_0(b), E^{(k)}_1(b), \dots $ are all one-dimen\-sional;
  \item 
    the operator $ H(b) $ 
    is unitarily equivalent to a direct-integral decomposition in the sense that 
    \begin{equation}\label{eq:unitary}
      \mathcal{F} H(b)\, \mathcal{F}^{-1} = \int_\mathbbm{R}^\oplus dk \, H^{(k)}(b).
    \end{equation} 
    Its spectrum $ \sigma(H(b)) $ is the set-theoretic union of 
    \emph{energy bands} defined as the closed intervals
    \begin{equation}\label{eq:band}
      \beta_n :=  
      \overline{\varepsilon^{(\mathbbm{R})}_n(b)} 
      = \overline{\big] \, \inf_{k \in \mathbb{R}} \varepsilon^{(k)}_n(b) , 
          \, \sup_{k \in \mathbb{R}} \, \varepsilon^{(k)}_n(b) \big[}
      \subseteq [0, \infty[, 
      \qquad n \in \mathbb{N}_0.
    \end{equation} 
    It has an absolutely continuous part $ \sigma_{\rm ac}(H(b)) = \bigcup_{|\beta_n| > 0 } \beta_n $ and a pure-point part 
    $ \sigma_{\rm pp}(H(b)) = \bigcup_{|\beta_n| = 0 } \beta_n $, the latter of which consists at most of infinitely degenerate eigenvalues. 
    The corresponding spectral projections $E_{\rm ac}(b)$ and $E_{\rm pp}(b)$ satisfy
    \begin{equation*}
      \mathcal{F} E_{\rm ac}(b) \, \mathcal{F}^{-1} = \! \sum_{|\beta_n|> 0} \int_\mathbbm{R}^\oplus\! dk \,  E^{(k)}_n(b), \qquad
      \mathcal{F} E_{\rm pp}(b) \, \mathcal{F}^{-1} = \! \sum_{|\beta_n|= 0} \int_\mathbbm{R}^\oplus\! dk \, E^{(k)}_n(b). 
    \end{equation*}
    {\rm [}Here and in the following $ | \cdot | $ denotes the one-dimensional Lebesgue measure.{\rm]}
  \end{nummer}
\end{proposition}
\begin{remarks}
  \begin{nummer}
     \item
       That the singular continuous spectrum of $ H(b) $  
       is empty, $ \sigma(H(b)) $\hspace{0pt}$= \sigma_{\rm ac}(H(b)) \cup \sigma_{\rm pp}(H(b)) $, also 
       follows from a rather general result on analytically fibered operators \cite{GeNi98}.
     \item
       Proposition~\ref{prop:spectral} assures that the $ n $th \emph{energy-band function} 
       $ \varepsilon_n(b): \mathbbm{R} \to \mathbbm{R}$, $k \mapsto  \varepsilon^{(k)}_n(b) $ is analytic for every 
       \emph{band index} $ n \in \mathbbm{N}_0 $. 
       If $ \varepsilon_n(b) $ is constant, equivalently, 
       if the \emph{bandwidth} $ | \beta_n | $ is zero, the $ n$th band $ \beta_n $ is 
       called \emph{flat}.  
       Because of the analyticity of $ \varepsilon_n(b) $, 
       the condition of a non-zero bandwidth, $ |\beta_n| > 0 $, is equivalent
       to  
       \begin{equation}\label{eq:bandflach}
         \left| \left\{ k \in \mathbb{R} \, : \, \frac{d \varepsilon^{(k)}_n(b)}{d k}= 0 \right\} \right| = 0.
       \end{equation} 
       Moreover, for all $ n \in \mathbb{N}_0 $ and all $ k \in \mathbb{R} $ one has the strict inequality 
       \begin{equation}\label{eq:group}
         \left(\frac{d \varepsilon^{(k)}_n(b)}{d k}\right)^{\!\! 2} <  2 \, \varepsilon^{(k)}_n(b).
       \end{equation}
       It is a consequence of 
       the Feynman-Hellmann formula (\cite[Ch. VII, \S 3.4]{Kat66} or \cite{IsZh88}) 
       \begin{equation}\label{eq:FH}
         \frac{d \varepsilon^{(k)}_n(b)}{d k} \, E^{(k)}_n(b) =   E^{(k)}_n(b) \left( k -  a(Q_1) \right)  E^{(k)}_n(b),
       \end{equation}
       the inequalities 
       \begin{equation*}
         \big( E^{(k)}_n(b) \left( k -  a(Q_1) \right)  E^{(k)}_n(b) \big)^2 \leq 
         E^{(k)}_n(b) \big( k -  a(Q_1) \big)^2  E^{(k)}_n(b) \leq 2 H^{(k)}(b) \,  E^{(k)}_n(b)
       \end{equation*}
	and the fact that $ \| P_1 \varphi \| > 0 $ for all $ \varphi \in \mathcal{D}\mathrm{om}(P_1) \backslash \{0\} $.	
     \item
       By~(\ref{eq:anti}) the \emph{effective scalar potential} 
       \begin{equation}\label{eq:effpot}
         v^{(k)} : \mathbbm{R} \to \mathbbm{R},\;  x_1 \mapsto v^{(k)}(x_1):=\frac{1}{2}( k - a(x_1) )^2,    
       \end{equation} 
       entering $ H^{(k)}(b) $ grows near infinity not slower than quadratically for any $ k \in \mathbbm{R} $. 
  \end{nummer}
\end{remarks}
\begin{proof}[Proof of Proposition~\ref{prop:spectral}]
   By checking the requirements of \cite[Def.~on p.~16]{ReSi4} the first assertion in part (i) follows from arguments along the lines of \cite[Lemma~2.4(b)]{Iwa85}. 
   By the (at least) quadratic growth of $ v^{(k)}$, the associated effective Hamiltonian 
   $ H^{(k)}(b) $ has only discrete spectrum \cite[Thm.~XIII.16]{ReSi4} with non-degenerate 
   eigenvalues $ (\varepsilon^{(k)}_n(b) )_{n \in \mathbbm{N}_0} $ \cite[Cor.~III.1.5]{CaLa90}. 
   Their analyticity as functions of $ k $ follows in turn from the fact that the family $ \left\{ H^{(k)}(b) \right\}_{k \in \mathbbm{R}} $ 
   is analytic of type A (cf.~\cite[Thm.~XII.8]{ReSi4}). 
   The unitary equivalence (\ref{eq:unitary}) derives from the identity 
   $  \mathcal{F} H(b) \psi = \int_\mathbbm{R}^\oplus dk \, H^{(k)}(b) \, \mathcal{F}  \psi $
   for all $ \psi \in \mathcal{D}\mathrm{om}(H(b)) $, the domain of $ H(b) $. 
   This is easily checked for $ \psi \in \mathcal{S}(\mathbbm{R}^2) $ and then follows for general $ \psi \in \mathcal{D}\mathrm{om}(H(b)) $
   from the essential self-adjointness of $ H(b) $ and $ H^{(k)}(b) $ 
   on $ \mathcal{S}(\mathbbm{R}^2) $, respectively, $ \mathcal{S}(\mathbbm{R}) $. 
   The condition of a non-zero bandwidth, $ |\beta_n| > 0 $, and hence (\ref{eq:bandflach}) 
   implies (cf. \cite[Thm.~XIII.86]{ReSi4} and \cite[Lemma~2.6]{Iwa85}) that the $ n $th band contributes to 
   the absolutely continuous spectrum of $ H(b) $. 
   In the other case, $ |\beta_n | = 0 $,  the $ n $th band contributes to the pure-point spectrum of $ H(b) $ \cite[Thm.~XIII.85]{ReSi4}. 
   The continuity of $ \varepsilon_n(b) $ guarantees the equality in (\ref{eq:band}).
   We finally note that the set-theoretic unions 
   $ \bigcup_{|\beta_n| > 0 } \beta_n $ and $ \bigcup_{|\beta_n| = 0 } \beta_n $ are closed sets, 
   since $ \sup_{k \in \mathbbm{R}} \varepsilon^{(k)}_n(b)  \subseteq [0,\infty]$ grows unboundedly as $ n \to \infty $. 
   This follows from the quadratic growth of $ v^{(k)}$ which implies the existence of 
   two constants $ \alpha>0 $ and $\gamma \in \mathbbm{R} $ 
   such that $ \alpha n + \gamma \leq \varepsilon^{(0)}_n(b) 
   \leq \sup_{k \in \mathbbm{R}} \varepsilon^{(k)}_n(b) $ for all 
   $ n \in \mathbbm{N}_0 $, by the min-max principle \cite{ReSi4}. 
\end{proof}

As already pointed out in Section~\ref{Intro}, there is the conjecture, which basically goes back to Iwatsuka, that 
there are no bounds states, $ E_{\rm pp}(b) = 0 $, (equivalently, $ \sigma_{\rm pp}(H(b))= \emptyset $, or 
$ | \beta_n | > 0 $ for all $ n \in \mathbbm{N}_0 $) holds true for 
general UMF's provided they are not globally constant \cite{CyFr87,MaPu97}.  
\begin{example}[Globally constant magnetic field]\label{ex:const}
  If $ b(x_1) = b_0 $  for Lebesgue-almost all $ x_1 \in \mathbbm{R} $ with a constant $ b_0 \in \mathbbm{R} \setminus \{0\} $, 
  one has a UMF with $ \overline{b} = | b_0 | $ and the Hamiltonian $ H^{(k)}(b) $ is 
  that of a displaced harmonic oscillator with $ k $-independent
  eigenvalues, $ \varepsilon^{(k)}_n(b) = (n + 1/2 ) \, | b_0 | $, $ n \in \mathbbm{N}_0 $. Consequently,
  the spectrum of $ H(b) $ is only pure-point and consists of infinitely degenerate, equidistant eigenvalues, 
  the well-known \emph{Landau levels} \cite{Foc28,Lan30}.
\end{example} 
Because of the analyticity of the eigenvalues $\varepsilon^{(k)}_n(b)$, a proof of Iwatsuka's conjecture amounts to rule out flat bands 
as they occur in the globally constant case, 
that is, to
prove (\ref{eq:bandflach}) for all $ n \in \mathbbm{N}_0 $. 
For Hamiltonians on $\mathrm{L}^2(\mathbbm{R}^d)$ with (rather general) $ \mathbbm{Z}^d $-periodic scalar potentials (only), 
the non-existence of 
flat bands 
has been proven several decades ago \cite{Tho73,ReSi4,Wil78,DyPe82}. 
One class of UMF's, for which (\ref{eq:bandflach}) was proven for all $ n \in \mathbbm{N}_0 $, concerns certain UMF's of a definite sign 
and is due to Iwatsuka himself.  
\begin{example}[Iwatsuka \cite{Iwa85}]\label{ex:iwa}
        Suppose that a UMF is (smooth,) strictly positive and bounded, that is, $ b_- \leq b(x_1) \leq b_+ < \infty $ 
        for Lebesgue-almost all $ x_1 \in \mathbbm{R} $ with some constants $ b_\pm > 0 $. 
        If additionally either $ \limsup_{x_1 \to \infty } b(x_1) < \liminf_{x_1 \to - \infty } b(x_1) $ or 
        $ \limsup_{x_1 \to -\infty } $ $b(x_1) < \liminf_{x_1 \to \infty } b(x_1) $, then $ | \beta_n | > 0 $ 
        for all $ n \in \mathbbm{N}_0 $ and hence the spectrum of $ H(b) $ is only absolutely continuous.
\end{example}
Another class of UMF's yielding only absolutely continuous spectrum of $ H(b) $  
covers in particular the UMF's of indefinite sign studied in \cite{Mul92} and \cite{RePe00}.
\begin{example}[Semi-bounded vector potential]
  Suppose that $ b $ is a UMF and that additionally its anti-derivative $ a $ 
  is globally bounded either from above or from below. 
  Then $ k_0 - a(x_1 ) $ has a definite sign for all $ x_1 \in \mathbbm{R} $ for a suitable $ k_0 \in \mathbbm{R} $. 
  By the Feynman-Hellmann formula (\ref{eq:FH}) and the unique-continuation property of eigenfunctions of 
  Schr\"o\-dinger operators \cite{ReSi4} one has $  d \varepsilon^{(k_0)}_n(b)/ d k_0 \neq 0 $ and hence $ |\beta_n | > 0 $ 
  for all $ n \in \mathbbm{N}_0 $.
  Therefore
  the spectrum of $ H(b) $ is only absolutely continuous. 
\end{example}
For yet another example, see \cite{MaPu97}. We stress that neither of these examples 
cover the typical realisations of UMF's being random in the sense of Section~\ref{sec:random} below.\\
%

In the following theorem we prove the continuity of 
the eigenvalues $\varepsilon^{(k)}_n(b) $, $ n \in \mathbbm{N}_0 $, 
of each effective Hamiltonian $ H^{(k)}(b) $ as a functional of $ b $ in case the latter has a definite sign. 
As in Example~\ref{ex:iwa}, it suffices to consider strictly positive UMF's.
The chosen distance 
\begin{equation}\label{eq:metric}
         \mathrm{d}(b, b') : = \sum_{j \in \mathbbm{Z}} 2^{-|j|}  \, \min\Big\{ 1 , \int_j^{j+1} dx_1 \, \big| b(x_1) -  b'(x_1) \big| \Big\}
\end{equation}
between two UMF's $ b $ and $ b' $ probes their absolute difference  only locally as 
given by the ${\rm L}^1_{\rm loc}(\mathbbm{R}) $-norm.
We will make use of the theorem in Section~\ref{sec:random}.
\begin{theorem}[Continuity of the eigenvalues at sign-definite UMF's]\label{thm:loccont}
  Let $ b $ and $ b_m $ for each $ m \in \mathbbm{N} $ be UMF's. 
  Suppose there exists a constant $ b_- \in ]0,\infty[ $ such that 
  the Lebesgue-essential ranges of $ b $ and $ b_m $ satisfy
  $  b(\mathbbm{R}) \subseteq [b_-, \infty[ $ and $ b_m(\mathbbm{R}) \subseteq [b_-, \infty[ $ for all $ m \in \mathbbm{N} $. 
  Then 
  \begin{nummer}
     \item
      $ \varepsilon^{(k)}_n(b) \in [(n+1/2) \, b_-, \infty[ \,$;
    \item
      the convergence $ \lim_{m \to \infty} \mathrm{d}(b_m, b) = 0 $
      implies the convergence
      \begin{equation}\label{eq:asstheorem}
        \lim_{m \to \infty}   \varepsilon^{(k)}_n(b_m) = \varepsilon^{(k)}_n(b)
      \end{equation}
  \end{nummer}
  for any band index $ n \in \mathbbm{N}_0 $ 
      and any wave number $ k \in \mathbbm{R} $.
\end{theorem}
\begin{remark}\label{rem:conv}
  Elementary arguments yield the inequalities
  \begin{equation}\label{eq:assumpconv}
    2^{-(\ell+1)} \min\Big\{1 , \int_{- \ell}^\ell dx_1 \, \big| b(x_1) \big| \Big\} \leq 
    \mathrm{d}(b, 0)
    \leq \int_{- \ell}^\ell dx_1 \, \big| b(x_1)  \big| + \sum_{|j| \geq \ell - 1} 2^{-|j|},
  \end{equation}
  valid for all real $ \ell > 0 $ and all $ b \in {\rm L}^1_{\rm loc}(\mathbbm{R}) $. 
  Hence $ \lim_{m \to \infty} \mathrm{d}(b_m, b) = 0 $ is equivalent to 
  $ \lim_{m \to \infty} \int_{- \ell}^\ell dx_1 \, \big| b_m(x_1) - b(x_1) \big| = 0 $
  for all $ \ell > 0 $.
\end{remark}
\begin{proof}[Proof of Theorem~\ref{thm:loccont}]
        Assertion~(i) follows from the first inequality in (\ref{eq:untereschrv}) below,
        the min-max principle \cite{ReSi4} and Example~\ref{ex:const}.
        For a proof of assertion~(ii) we fix $ k \in \mathbbm{R} $ and let $ \xi^{(k)}_m \in \mathbbm{R} $ 
        denote, for each $ m \in \mathbbm{N} $, the solution of the equation $  a_m(\xi^{(k)}_m) = k $, which is unique because
        the (absolutely) continuous function 
        $ x_1 \mapsto a_m(x_1) = \int_0^{x_1}dy_1 \, b_m(y_1) $ is strictly increasing. This solution obeys the estimate
        $  |\xi^{(k)}_m| \leq |k|/b_- $ for all $ m \in \mathbbm{N} $. 
        As a consequence, the effective potential (\ref{eq:effpot}) associated with 
        $ b_m $ is bounded from below by a quadratic potential according to
        \begin{equation}\label{eq:untereschrv}
        2 \, v^{(k)}_m(x_1) =  \Big( \int_{\xi^{(k)}_m}^{x_1} \mkern-5mu dy_1 \, b_m(y_1) \Big)^2 
        \geq  b_-^2  \left( x_1 - \xi^{(k)}_m \right)^2 
        \geq \frac{b_-^2}{2} \, x_1^2 - k^2
        \end{equation}
        for all $ x_1 \in \mathbbm{R} $. Therefore the shifted effective Hamiltonian 
        $ H^{(k)}(b_m) + k^2/2  $ is bounded from below by the self-adjoint harmonic-oscillator Hamiltonian  
        $ H_0 := P_1^2/2 + b_-^2 \, Q_1^2/4  $ on $ {\rm L}^2(\mathbbm{R}) $. 
        Hence one gets the resolvent estimate $ R^{(k)}(b_m):= (H^{(k)}(b_m) + k^2/2)^{-1} \leq H_0^{-1} $ for all $ m \in \mathbbm{N} $ 
        by the operator monotonicity of the reciprocal function (cf.\ \cite[Prop.~A.2.5]{GlJa87}).
        The same lines of reasoning imply $ R^{(k)}(b) := (H^{(k)}(b) +  k^2/2)^{-1} \leq H_0^{-1} $. 
        Since all involved resolvents are compact, the dominated-convergence theorem for 
        compact operators \cite[Thm.~2.16(b)]{Sim79a} ensures that the norm-resolvent convergence of $  H^{(k)}(b_m) $ to $ H^{(k)}(b) $ as
        $ m \to \infty $, that is
        \begin{equation}\label{eq:normres}
                \lim_{m \to \infty} \big\| R^{(k)}(b_m) -  R^{(k)}(b) \big\| = 0,      
        \end{equation}
        is implied by the respective strong-resolvent convergence. 
        Here, $ \| B \| := \sup_{\| \varphi \| = 1 } \| B \varphi \| $ is the usual norm of a bounded operator $ B $ on 
        ${\mathrm L}^2(\mathbbm{R}) $ where the supremum is taken over all normalised
        $ \varphi \in {\mathrm L}^2(\mathbbm{R}) $.
        Now, to prove strong-resolvent convergence 
        it suffices \cite[Thm.~VIII.25]{ReSi1}  to show that 
        \begin{equation}
                \lim_{m \to \infty} \left\| H^{(k)}(b_m)\, \varphi - H^{(k)}(b)\, \varphi \right\|^2 =        
                \lim_{m \to \infty}
         \int_{\mathbbm{R}} dx_1 \, \big| v^{(k)}_m(x_1) - v^{(k)}(x_1) \big|^2 \, |\varphi(x_1)|^2  = 0  
        \end{equation}
        for all $ \varphi \in \mathcal{C}_0^\infty(\mathbbm{R}) $, the space of arbitrarily often differentiable and 
        compactly supported functions, because the effective Hamiltonians are 
        essentially self-adjoint on $ \mathcal{C}_0^\infty(\mathbbm{R}) $ \cite[Thm.~X.28]{ReSi2}. 
        In fact, the last equality follows from $ \lim_{m \to \infty} \mathrm{d}(b_m, b) = 0 $, Remark~\ref{rem:conv} and the estimate
        \begin{align}
                & 2 \, \sup_{x_1 \in [-\ell ,\ell]}  \big| v^{(k)}_m(x_1) - v^{(k)}(x_1) \big| \notag \\ 
                & = 
                \sup_{x_1 \in [-\ell ,\ell]} \big| 
                \big( a_m(x_1) - a(x_1) \big) \big[ a_m(x_1) - a(x_1) - 2 (k - a(x_1) ) \big]  \big| \notag \\
                & \leq \big\| b_m - b \big\|_{1, \ell} \left[ \big\| b_m - b \big\|_{1, \ell} + 2 
                  \big( |k| + \big\| b \big\|_{1, \ell} \big) \right]
        \end{align}  
        which is valid for all real 
        $ \ell > 0 $ and relies on the inequality $ \sup_{x_1 \in [-\ell ,\ell]} | a_m(x_1) - a(x_1) | \leq  \int_{-\ell}^\ell  
                dx_1 \, \big| b_m(x_1) - b(x_1) \big| =: \big\| b_m - b \big\|_{1, \ell} $. 
         This completes the proof of (\ref{eq:normres}). The claimed convergence 
        (\ref{eq:asstheorem}) of the eigenvalues eventually follows therefrom and from the inequality  
        \begin{equation}
          \big| \big(\varepsilon^{(k)}_n(b_m) +   k^2/2\big)^{-1} - \big(\varepsilon^{(k)}_n(b) +   k^2/2\big)^{-1} \big|
        \leq  \big\| R^{(k)}(b_m) -  R^{(k)}(b) \big\|,
        \end{equation}
        which is 
        valid for all $ n \in \mathbbm{N}_0 $ and all $ m \in \mathbbm{N} $ \cite[Prob.\ 2 on p.\ 364]{ReSi4}. 
\end{proof}
\subsection{Energy bands and some transport properties}
Since the magnetic field depends anisotro\-pically on the two coordinates, any normalised wave packet $ \psi_0 \in {\rm L}^2(\mathbb{R}^2) $, $\|\psi_0\|=1$, 
which is initially localised along one direction, should expand
anisotropically over the plane under its time evolution $ \psi_t := e^{ -i t H(b) } \, \psi_0 $, $t \in \mathbbm{R}$, generated by (\ref{eq:H}).  
As a simple degree for the expansion 
along the $ x_j $-direction ($ j \in \{ 1, 2 \} $) 
we use the corresponding second spatial moment
\begin{equation}
  \| Q_j  \psi_t \|^2  = \int_{\mathbb{R}^2} \mathrm{d}^2 x \, | \psi_t(x) |^2 \, x_j^2
\end{equation}
 of the (pure) quantum state given by $ \psi_t \in \mathcal{D}\mathrm{om}(Q_j) $ 
in the (maximal) domain of $ Q_j $.   
By switching to the Heisenberg picture it can also be written as $ \| Q_{j,t}   \psi_0 \|^2 $ 
in terms of the time-evolved position operator
$ Q_{j, t} := e^{i t H(b) } Q_j \, e^{- i t H(b) } $.
Our first result on the quantum dynamics is simple. Due to the (at least) quadratic confinement of the particle 
by the effective scalar potential 
for large $ | x_1 | $, wave packets do not spread along the $ x_1 $-direction in the course of time.
\begin{theorem}[Dynamical localisation along the $ x_1 $-direction]\label{thm:dynloc}
  Let $ b $ be a UMF. Then any normalised 
  wave packet with finite total kinetic energy, $ \psi_0 \in \mathcal{D}\mathrm{om}(H(b)^{1/2}) $, 
  which is initially localised in
  the sense that $ \psi_0 \in \mathcal{D}\mathrm{om}( Q_1 ) $ and
  $ \psi_0 \in \mathcal{D}\mathrm{om}(a(Q_1)) $, remains localised for all times, 
\begin{equation}\label{eq:dynloc}
  \sup_{t \in \mathbb{R}} \;  \| Q_{1} \,  \psi_t \|  < \infty.
\end{equation}
\end{theorem}
\begin{remarks}
  \begin{nummer}
  \item
    The two initial-localisation conditions are fulfilled for any $ \psi_0 \in \mathcal{S}(\mathbbm{R}^2) $. 
    For more general $ \psi_0 \in {\rm L}^2(\mathbbm{R}^2) $, the first condition, $ \| Q_1 \,  \psi_0 \| < \infty $, implies the 
    second one, $ \| a(Q_1) \psi_0 \| < \infty $, if $ \lim_{|x_1| \to \infty } |a(x_1)|/|x_1| = \overline{b} > 0 $ 
    (as will be the case by ergodicity for a UMF being random in the sense of Section~\ref{sec:random}).
  \item
    For the validity of (\ref{eq:dynloc}) the requirement $ \overline{b} > 0 $ (in Definition~\ref{ass}) cannot simply be dispensed with. 
    For example, if a given absolutely continuous function $ a : \mathbbm{R} \to \mathbbm{R} $ is $\mathbbm{Z}$-periodic, one has
    $ \overline{b} = 0 $ and 
    the corresponding Hamiltonian (\ref{eq:H}) on $ {\rm L}^2(\mathbbm{R}^2) $
    also fibres into a one-parameter family of effective Hamiltonians 
    $  \left\{ H^{(k)}(b) \right\}_{k \in \mathbbm{R}} $ on $ {\rm L}^2(\mathbbm{R}) $,
    but each member of which is $ \mathbbm{Z} $-periodic and hence has only absolutely continuous spectrum \cite{Tho73,ReSi4,Wil78}.  
    The dynamical characterisation of scattering states in Hilbert space by the RAGE-theorem \cite{CyFr87,Weid03} 
    therefore implies (for the present situation of one dimension and without singular continuous spectrum)
    the second of the following two equalities 
    \begin{equation}\label{eq:nolocexample}
       \lim_{t \to \infty} \left\| \chi_{[-r,r]} (Q_1) \, e^{-itH(b)} \psi_0 \right\|^2 
       = \int_{\mathbbm{R}}\mkern-5mu dk  
       \lim_{t \to \infty} \left\| \chi_{[-r,r]} (Q_1) \, e^{-itH^{(k)}(b)} \left(\mathcal{F} \psi_0 \right)^{(k)} \right\|^2 = 0 
    \end{equation}  
    for any real $ r > 0 $, where $ x_1 \mapsto  \chi_{[-r,r]}(x_1) $ denotes the indicator function of the interval $ [-r, r] $. 
    The first equality in (\ref{eq:nolocexample}) is due to the dominated-convergence theorem and 
    the fact that the partial Fourier transformation (\ref{eq:ParFour}) is an isometry 
    which commutes with $ Q_1 $.  
    Since $ x_1^2 \geq r^2 \left( 1 - \chi_{[-r,r]}(x_1) \right) $ for all $ x_1 \in \mathbb{R} $ and hence 
    $ \| Q_{1} \,  \psi_t \|^2 \geq r^2 \left( 1 - \| \chi_{[-r,r]} (Q_1) \psi_t \|^2 \right) $ for any (arbitrarily large) $ r > 0 $, 
    Eq.~(\ref{eq:nolocexample}) implies that $ \| Q_1 \, \psi_t \| $ grows unboundedly with increasing $ t $ for these examples of 
    $\mathbbm{Z} $-periodic $ b \in {\mathrm L}^1_{\mathrm{loc}}(\mathbbm{R}) $ defined by $ b(x_1) := da(x_1)/ dx_1 $
    (for Lebesgue-almost all $ x_1 \in \mathbbm{R} $).
\end{nummer}  
\end{remarks}
\begin{proof}[Proof of Theorem~\ref{thm:dynloc}]
According to Assumption~\ref{ass}, there exists a length scale $ r > 0 $ 
such that $ \overline{b}\, | x_1 | / 2 \leq |a(x_1)|  $ 
for all $ x_1 \in \mathbbm{R} $ with $ |x_1| > r $. 
As a consequence, we have $ | x_1 | \leq r + \big( 2/ \overline{b} \big) \, | a(x_1) | $ for all
$ x_1 \in \mathbbm{R} $ and therefore 
\begin{equation}\label{eq:q1}
  \big\| Q_1 \psi_t \big\| \leq r + \left(2/ \overline{b} \, \right) \, 
  \big\| a(Q_1) \psi_t \big\|.
\end{equation}
Using the inequality
\begin{equation}\label{eq:aq1}
  \big( \| P_2 \psi_0 \| - \|  a(Q_1) \psi_s \| \big)^2  \leq 2 \, \| H(b)^{1/2} \, \psi_0 \|^2,
\end{equation}
being valid for all $ s \in \mathbb{R}$, first for $ s = t $ and then for
$ s = 0 $ we bound the second term
on the right-hand side of (\ref{eq:q1}) by a time-independent one according to
$  \| a(Q_1) \psi_t \| \leq 2 \sqrt{2} \| H(b)^{1/2} \, \psi_0 \| +  \| a(Q_1) \psi_0 \| $.
The validity of (\ref{eq:aq1}) itself follows from the
triangle inequality and the fact that $ P_2 $ and $ H(b) $ are 
constants of the motion, that is, commute with $ H(b) $.
\end{proof}

For a description of the long-time behaviour along the $ x_2 $-direction, we introduce an operator 
$ \overline{V}_{2,\infty} := \mathcal{F}^{-1}\int_{\mathbbm{R}}^\oplus dk \, \overline{V}_{2,\infty}^{(k)}  \, \mathcal{F} $ on $  \mathcal{D}\mathrm{om}(H(b)^{1/2}) $  in terms of its fibres 
\begin{equation}\label{eq:asymv}
  \overline{V}_{2,\infty}^{(k)} := \sum_{n=0}^\infty \frac{d \varepsilon^{(k)}_n(b)}{d k} \, E^{(k)}_n(b), \quad k \in \mathbb{R},
\end{equation}  
on  $ \mathcal{D}\mathrm{om}(H^{(k)}(b)^{1/2}) $. 
Our next task is to show that $ \overline{V}_{2,\infty} $  is the asymptotic velocity operator (in the sense of \cite{DeGe97}) corresponding 
to the motion along the $ x_2 $-direction. To do so, we first make sure 
that $ \overline{V}_{2,\infty} $ is well-defined and collect some of its properties.
\begin{lemma}[Properties of the asymptotic velocity]\label{lemma:asvel}
  Let $ b $ be a UMF. Then the operator $ \overline{V}_{2,\infty} $ is bounded from $ \mathcal{D}\mathrm{om}(H(b)^{1/2})$ to $ {\rm L}^2(\mathbbm{R}^2) $ according to
  \begin{equation}\label{eq:asymkin}
    \| \overline{V}_{2,\infty} \psi \| < \sqrt{2} \, \| H(b)^{1/2} \psi \| 
  \end{equation}
  for all $ \psi \in  \mathcal{D}\mathrm{om}(H(b)^{1/2}) $. 
  Moreover, one has:
  \begin{enumerate}
    \item[(i)]
      $ \overline{V}_{2,\infty} E_{\rm ac}(b) = \overline{V}_{2,\infty} $ and $ \| \overline{V}_{2,\infty} \psi \| > 0 $ for all $ \psi \in E_{\rm ac}(b) \, \mathcal{D}\mathrm{om}(H(b)^{1/2}) $;
    \item[(ii)]
      $ \overline{V}_{2,\infty} E_{\rm pp}(b) = 0 $.
  \end{enumerate}  
\end{lemma}
\begin{remark} 
  The relation of the asymptotic velocity operator to the energy-band functions is similar to that 
  for one-dimensional motion in a $\mathbbm{Z}$-periodic scalar potential \cite{GeNi98b,AsKn98}. 
  In case of a globally constant magnetic field (cf. Example~\ref{ex:const}), for which $ E_{\rm ac}(b) = 0 $, 
  the asymptotic velocity vanishes, $  \overline{V}_{2,\infty} = 0 $, in accordance with physical intuition.
  In any case, the strict inequality (\ref{eq:asymkin}) simply means that the asymptotic kinetic energy of the particle's motion 
  along the $x_2$-direction is always smaller than its (time-invariant) total kinetic energy; cf.\ Theorem \ref{thm:ball} below.
\end{remark}
\begin{proof}[Proof of Lemma~\ref{lemma:asvel}] 
  The proof of (\ref{eq:asymkin}) is based on (\ref{eq:sectral}) and (\ref{eq:group}) which yield 
  \begin{equation}
    \left\| \overline{V}_{2,\infty}^{(k)} \varphi \right\|^2 = \sum_{n=0}^\infty \Bigg(\frac{d \varepsilon^{(k)}_n(b)}{d k}\Bigg)^2 
    \left\|   E^{(k)}_n(b) \varphi \right\|^2 <
  2 \, \big\| H^{(k)}(b)^{1/2} \varphi \big\|^2 
  \end{equation}
  for all $ \varphi \in \mathcal{D}\mathrm{om}(H^{(k)}(b)^{1/2}) $.
  Since the partial Fourier transformation (\ref{eq:ParFour}) is an isometry, one therefore has
  \begin{align}\label{eq:gerechne}
    \left\| \overline{V}_{2,\infty} \psi \right\|^2 &=  \int_\mathbbm{R} \!\! dk \, 
    \left\| \overline{V}_{2,\infty}^{(k)} \left(\mathcal{F}\psi\right)^{(k)} \right\|^2 \notag \\
    & <  2 \int_\mathbbm{R} \!\! dk \, \left\| H^{(k)}(b)^{1/2} \left(\mathcal{F}\psi\right)^{(k)} \right\|^2 = 
     2 \left\| H(b)^{1/2} \psi \right\|^2 
  \end{align}
  for all $ \psi \in \mathcal{D}\mathrm{om}(H(b)^{1/2}) $. 
  For a proof of assertions~(i) and (ii) we note that only those terms contribute to the series in (\ref{eq:asymv}) 
  for which $ |\beta_n | > 0 $. 
  Thanks to the analyticity of $ \varepsilon^{(k)}_n(b) $ the latter is the case if and only if (\ref{eq:bandflach}) holds, 
  which implies that $\|  \overline{V}_{2,\infty}^{(k)} E^{(k)}_n(b) \varphi \| > 0 $
  for all $ \varphi \in  E^{(k)}_n(b) \, \mathcal{D}\mathrm{om}(H^{(k)}(b)^{1/2}) $ and Lebesgue-almost all $ k \in \mathbbm{R} $. The second assertion in (i) is thus proven with the help 
  of the first equality in (\ref{eq:gerechne}).
\end{proof}
We are now prepared to present our second result on the quantum dynamics. 
It concerns the long-time limit of the 
motion along the $ x_2 $-direction and, after all, justifies the name ``asymptotic velocity operator'' for $ \overline{V}_{2,\infty} $.
\begin{theorem}[Ballistic transport along the $ x_2 $-direction in the absence of flat bands]\label{thm:ball}
  Let $ b $ be a UMF. Then any normalised wave packet with finite total kinetic energy, $ \psi_0 \in \mathcal{D}\mathrm{om}(H(b)^{1/2}) $, 
  and initial localisation in the sense that $ \psi_0 \in \mathcal{D}\mathrm{om}(Q_2 )$, has $ \overline{V}_{2,\infty} $ as its 
  asymptotic velocity operator in the following limiting sense 
  \begin{equation}\label{eq:asymvel}
    \lim_{t \to \infty} \;  \left\| \frac{Q_{2,t} \,  \psi_0}{t} - \overline{V}_{2,\infty} \, \psi_0 \right\| = 0 .
  \end{equation}
  If additionally the entire spectrum of $ H(b) $ is absolutely continuous, equivalently
  $ | \beta_n | >  0 $ for all $ n \in \mathbbm{N}_0 $, 
  the motion is ballistic in the sense that 
  $0 < \|\overline{V}_{2,\infty} \,  \psi_0 \| < \infty $. 
\end{theorem}
\begin{remark}
  Eq.~(\ref{eq:asymvel}) implies $ \lim_{t \to \infty} f(Q_{2,t}/t) \, \psi = f(\overline{V}_{2,\infty}) \, \psi $ for
  all bounded and continuous functions $ f : \mathbbm{R} \to \mathbbm{R} $ and all $ \psi \in {\rm L}^2(\mathbbm{R}^2) $, a result which was
  already proven \cite[Thm.~4.2]{MaPu97} for certain UMF's. 
  Here we give an argument for the validity of the slightly stronger assertion (\ref{eq:asymvel}), 
  which closely follows the lines of reasoning of~\cite[Thm.~2.3]{AsKn98}.
\end{remark}
\begin{proof}[Proof of Theorem~\ref{thm:ball}]
  We first introduce the \emph{time-averaged velocity operator}
  \begin{equation}
          \overline{V}_{2,t}  :=  \frac{1}{t} \int_0^t \! d s \; e^{i s H(b) }\big( P_2 -  a(Q_1) \big) \, e^{- i s H(b) }
         =   \mathcal{F}^{-1}\int_{\mathbbm{R}}^\oplus\! dk \, \overline{V}_{2,t}^{(k)} \, \mathcal{F}
  \end{equation}        
  which is defined for $ t \neq 0 $ on $ \mathcal{D}\mathrm{om}(H(b)^{1/2}) $ with its fibres 
  \begin{equation}\label{eq:spec}
    \overline{V}_{2,t}^{(k)} := \frac{1}{t} \int_0^t \! d s \; e^{i s H^{(k)}(b) }\big( k -  a(Q_1) \big) \, e^{- i s H^{(k)}(b) }
  \end{equation}
  on $ \mathcal{D}\mathrm{om}(H^{(k)}(b)^{1/2}) $. Since 
  \begin{equation}
    \big\| \overline{V}_{2,t}  \, \psi_0 \big\| 
    \leq  \frac{1}{t}  \int_0^t \! d s \, \Big\| \big( P_2 - a(Q_1) \big) e^{-isH(b)} \psi_0 \big\| < \sqrt{2}  \, \Big\| H(b)^{1/2} \psi_0 \Big\|, 
  \end{equation}
  $  \overline{V}_{2,t} $ is bounded from
  $ \mathcal{D}\mathrm{om}(H(b)^{1/2}) $ to $ {\rm L}^2(\mathbbm{R}^2) $, uniformly in $ t \in \mathbbm{R}\backslash \{0\} $.
  Arguments as in \cite[Thm.~2.1]{RaSi78} then justify that
  the time-evolved second component of the position operator 
  acts in the standard way,
  $    Q_{2,t}\, \psi_0  = Q_2 \, \psi_0 +  t\, \overline{V}_{2,t}  \, \psi_0 $
  on any (normalised) $ \psi_0 \in \mathcal{D}\mathrm{om}(H(b)^{1/2}) \cap \mathcal{D}\mathrm{om}(Q_2) $. The assertion~(\ref{eq:asymvel}) is thus equivalent to 
  \begin{equation}\label{eq:equiv}
    \lim_{t \to \infty} \left\| \overline{V}_{2,t} \psi_0 - \overline{V}_{2,\infty} \psi_0 \right\| = 0 
\end{equation}
for all $ \psi_0 \in \mathcal{D}\mathrm{om}(H(b)^{1/2}) $. By the uniform boundedness (in $ t \in \mathbbm{R}\backslash \{0\} $) of 
$ \overline{V}_{2,t} $ on the domain $ \mathcal{D}\mathrm{om}(H(b)^{1/2}) $ it suffices
to prove (\ref{eq:equiv}) for any $ \psi_0 $ in the finite-band-index subspace
\begin{equation}
  \mathcal{E}:=\left\{ \psi \in  {\rm L}^2(\mathbbm{R}^2) \, : \, \mathcal{F} \psi =   \sum_{n= 0}^l \int_\mathbbm{R}^\oplus dk \, E^{(k)}_n(b) \mathcal{F} \psi  \quad 
    \mbox{for some $ l \in \mathbbm{N}_0$} \right\}
\end{equation} 
which is dense in $ \mathcal{D}\mathrm{om}(H(b)^{1/2}) $.
Now, let $ \psi_0 \in \mathcal{E} $ arbitrary and $  l \in \mathbbm{N}_0 $ its maximal band index. Then the following equalities hold
\begin{align}\label{eq:velred}
   & \big\|\big( \overline{V}_{2,t} -\overline{V}_{2,\infty}\big) \psi_0 \big\|^2 
   = \int_\mathbbm{R} \! dk \, \big\| \big( \overline{V}_{2,t}^{(k)} -\overline{V}_{2,\infty}^{(k)}\big) \left(\mathcal{F}\psi_0\right)^{(k)} \big\|^2  = \\
   & \int_\mathbbm{R} \! dk \, \sum_{n = 0}^\infty \Big\| \sum_{\substack{m= 0\\ m\neq n}}^l \frac{1}{t} \int_0^t \! d s \, 
   e^{i s \big(\varepsilon^{(k)}_n(b) - \varepsilon^{(k)}_m(b)\big)} \, E^{(k)}_n(b) \,
   \big(k - a(Q_1) \big)\, E^{(k)}_m(b) \left(\mathcal{F}\psi_0\right)^{(k)}  \Big\|^2. \notag 
\end{align}
The second equality derives from (\ref{eq:FH}) and (\ref{eq:spec}). The convergence (\ref{eq:equiv}) for $ \psi_0 \in \mathcal{E} $  
now follows from the fact that 
$ \lim_{t \to \infty} t^{-1} \int_0^t ds \, \exp\left\{i s (\varepsilon^{(k)}_n(b) - \varepsilon^{(k)}_m(b)\right\}  = 0 $ if $ m \neq n $ 
together with the dominated-convergence theorem. The latter is applicable since the squared norm on the right-hand side of (\ref{eq:velred}) has the upper bound 
\begin{equation}
    (l +1) \, \max_{j \in \{0, \dots , l\}} \big\| E^{(k)}_n(b) 
    \left(k - a(Q_1) \right)  E^{(k)}_j(b)  \left(\mathcal{F}\psi_0\right)^{(k)} \big\|^2,
\end{equation}
which is summable with respect to $ n\in \mathbbm{N}_0  $ and Lebesgue integrable with respect to $ k \in\mathbbm{R} $. This completes the proof of (\ref{eq:asymvel}). 
The assertion about ballistic transport in case $ E_{\rm pp}(b) = 0 $ follows from Lemma~\ref{lemma:asvel}.
\end{proof}


\section{Schr{\"o}dinger operators with random unidirectionally constant magnetic fields}\label{sec:random}

Throughout this section we are dealing with unidirectionally constant magnetic fields given by realisations 
$ b: \mathbbm{R} \to \mathbbm{R}  $  of 
an $ \mathbbm{R} $-valued random (or: stochastic) process with parameter set $ \mathbbm{R} $ in the sense of  
\begin{definition}[RUMF]\label{assr}
A \emph{random unidirectionally constant magnetic field} (RUMF) is a probability space 
$ (\Omega, \mathcal{B}(\Omega) , \mathbbm{P} ) $ with $ \Omega := \left\{ b \in {\mathrm L}^1_{\mathrm{loc}}(\mathbbm{R}) \, : \, 
  b(\mathbbm{R}) \subseteq \mathbbm{R} \right\} $
as its 
set of \emph{realisations} (or: sample paths) and with the collection $ \mathcal{B}(\Omega) $ of all Borel subsets of $ \Omega $ as its sigma-algebra 
of \emph{events}. 
The fixed measurable space $ (\Omega, \mathcal{B}(\Omega) ) $ is endowed with a 
probability measure $ \mathbbm{P} $ having two properties: 
\begin{enumerate} 
  \item[(i)] $ \mathbbm{P} $ is $ \mathbbm{R}$-ergodic; 
  \item[(ii)] $ \mathbbm{P} $ has a non-zero and finite mean-value, that is, $ 0 <| \int_\Omega \mathbbm{P}(db) \, b(x_1) | < \infty$ for 
    Lebesgue-almost all $ x_1 \in \mathbbm{R} $.
\end{enumerate}
\end{definition}
\begin{remarks}
  \begin{nummer}
    \item
      The metric $ \mathrm{d}: \Omega\times \Omega \to [0, 3] $ given by (\ref{eq:metric}) 
      renders $ \Omega $ a \emph{Polish space} (cf.\ \cite{Bau01}).
      The Borel sigma-algebra $ \mathcal{B}(\Omega) $ is the smallest sigma-algebra in $ \Omega $ 
      containing all subsets of $ \Omega $
      which are open with respect to $  \mathrm{d} $. 
      The \emph{topological support} of the probability measure in $ \Omega $ is the (closed) event
      \begin{equation}\label{eq:support}
        \mathrm{supp} \; \mathbbm{P} := \left\{ b \in \Omega \, : \, \mathbbm{P}\left( \Delta_\delta(b) \right) > 0
          \; \mbox{for all $ \delta > 0 $}\right\}, 
      \end{equation}
      where $ \Delta_\delta(b) := \left\{ b' \in \Omega \, : \, \mathrm{d}(b,b') < \delta \right\} $ is the open ball 
      with centre $ b \in \Omega $ and radius $ \delta > 0 $.  
    \item\label{rem:ergod}
      By defining $ (\theta_{z_1} b)(x_1) := b(x_1+z_1) $ for all $ z_1 \in \mathbbm{R} $, 
      Lebesgue-almost all $ x_1 \in \mathbbm{R} $ and any $ b \in \Omega $, one gets a group $\{\theta_{z_1} \}_{z_1 \in \mathbb{R}}$ of measurable \emph{shifts} 
      on $  (\Omega, \mathcal{B}(\Omega) ) $. 
      The probability measure $ \mathbbm{P} $ (and the resulting RUMF) 
      is \emph{$ \mathbbm{R} $-homogeneous} if $ \mathbbm{P}(\theta_{z_1} \Delta) = 
      \mathbbm{P}(\Delta)$ for all $ z_1 \in \mathbbm{R} $ and all 
      $ \Delta \in \mathcal{B}(\Omega) $. It is \emph{$ \mathbbm{R} $-ergodic} if, additionally, 
      every shift-invariant event $\Delta \in \mathcal{B}(\Omega) $, 
      $ \theta_{z_1} \Delta = \Delta $ for all $ z_1 \in \mathbbm{R} $, 
      is either \emph{almost impossible} or \emph{almost sure}, $ \mathbbm{P}(\Delta) \in \{0,1\} $.
    \item
      Due to the $\mathbbm{R} $-homogeneity of 
      $ \mathbbm{P} $ the (path) integral 
      for its mean-value\\ 
	$ \int_\Omega \mathbbm{P}(db) \, b(x_1) := (2 \ell)^{-1} \int_\Omega \mathbbm{P}(db)\int_{x_1-\ell}^{x_1 +\ell} dy_1\, b(y_1) $,
	with $ \ell> 0 $ arbitrary, 
       does not depend on 
      Lebesgue-almost all $ x_1 \in \mathbbm{R} $.
      In the following we adopt the convention to denote the corresponding constant by $\int_\Omega \mathbbm{P}(db) \, b(0) $.
    \item
      The probability measure of a RUMF can be specified by its 
      \emph{characteristic functional} given by $ \widetilde{\mathbbm{P}}(\eta):=
      \int_\Omega \mathbbm{P}(db) \, \exp\left\{ - i \int_{\mathbbm{R}} dx_1 \, \eta(x_1)\, b(x_1) \right\} $ 
      for all real-valued $ \eta \in \mathcal{C}_0^\infty(\mathbbm{R}) $, cf.\ \cite{Ito}.  
 \end{nummer}
\end{remarks}
As a first result, we show that $ \mathbbm{P}$-almost every realisation $ b: \mathbbm{R} \to \mathbbm{R} $, 
$ x_1 \mapsto b(x_1) $ of a RUMF is a UMF 
  in the sense of Definition~\ref{ass}. 
\begin{lemma}[Realisations of a RUMF are almost surely UMF's]
Let $ (\Omega, \mathcal{B}(\Omega) , \mathbbm{P} ) $ be a RUMF and define
$ \Omega_0 := \left\{ b \in \Omega \, : \,  b \; \mbox{is a UMF}\right\} $.
Then 
\begin{enumerate}
	\item[(i)] $ \Omega_0 $ is an almost-sure event, $ \mathbbm{P}(\Omega_0) = 1 $;
	\item[(ii)] for any $ b \in \Omega_0 $ the two constants associated with it according to 
	Definition~\ref{ass} are given by 
	$  \overline{b} =  |\int_\Omega \mathbbm{P}(db') \, b'(0)| $ and $ \alpha =  1 $.
\end{enumerate}
\end{lemma}
\begin{proof}
  We first note that $ \Omega_0 \in \mathcal{B}(\Omega) $, because the functional $ \Omega \ni b \mapsto a(x_1) = 
  \int_0^{x_1} dy_1 b(y_1) $ is measurable for every $ x_1 \in \mathbbm{R} $ such that the lower and upper limits in (\ref{eq:anti})
  are measurable functionals of $ b $. 
  In fact, taking there $\alpha = 1 $ these limits coincide with 
  $ \overline{b} =  \big|\int_\Omega \mathbbm{P}(db') \, b'(0)\big| > 0 $ for $ \mathbbm{P} $-almost all $b \in \Omega $, 
  since
  the Birkhoff-Khinchin ergodic theorem \cite{CrLe67,CoFoSi82,Kal02} yields the identity
  \begin{equation}\label{eq:mittel}
    \lim_{|\ell| \to \infty} \, \frac{1}{\ell } \, \int_{0}^{\ell } \!\! \mathrm{d}x_1 \; b(x_1) = 
    \int_\Omega\! \mathbbm{P}(db') \, b'(0)
  \end{equation}
  for $ \mathbbm{P}$-almost all $ b \in \Omega $. 
\end{proof}
\begin{remark}
  As a consequence, all results of Section 2 apply to every $ b\in \Omega_0 $, that is, to the RUMF-case with probability $ 1 $. 
  In particular, each realisation $ H^{(k)}(b) $ of any random effective Hamiltonian
  has  non-degenerate, strictly positive and isolated eigenvalues $ \varepsilon^{(k)}_n(b) $, $n \in \mathbbm{N}_0$. 
  For each fixed $ n$, they have two basic properties: (i)~
  the mapping $ \Omega_0 \times \mathbbm{R} \ni (b,k) \mapsto \varepsilon^{(k)}_n(b) $ 
  is  measurable (cf. \cite[Sec.~V.1]{CaLa90}), hence an $ \mathbbm{R} $-valued random process with parameter set $ \mathbbm{R} $,
  and ~(ii)~ 
  its realisation $ \mathbbm{R} \ni k \mapsto \varepsilon^{(k)}_n(b) $ has an analytic extension to some  complex neighbourhood
  of $ \mathbbm{R} $ for any $ b \in \Omega_0 $ (cf.\ Proposition~\ref{prop:spectral}). 
\end{remark}
\subsection{Non-randomness of the energy bands}
It is a comforting fact to learn that although the spectrum of $H^{(k)}(b) $ in general depends on $ b \in \Omega_0 $ for each fixed 
$ k \in \mathbbm{R} $, each resulting energy band
of $ H(b) $
(cf.\ Proposition~\ref{prop:spectral}) is the same for $ \mathbbm{P} $-almost all $ b \in \Omega_0 $. 
\begin{theorem}[Almost-sure non-randomness of the energy bands]\label{thm:nreb}
  Let $ (\Omega, \mathcal{B}(\Omega) , \mathbbm{P} ) $ be a RUMF. 
  Then there exists a sequence $ \beta := ( \beta_n )_{n \in \mathbbm{N}_0} $ of non-random closed intervals 
  $  \beta_n \subseteq [0,\infty[$  such that 
  \begin{nummer}
    \item
      the event
      \begin{equation}\label{eq:nreb}
        \Omega_\beta := \left\{ b \in \Omega_0 \, : \quad \overline{\varepsilon^{(\mathbbm{R})}_n(b) } = \beta_n   \quad 
          \mbox{for all $ n \in \mathbb{N}_0 $} \right\} 
      \end{equation}
      is almost sure, $ \mathbbm{P}(\Omega_\beta) = 1 $;
    \item
      each event
      \begin{equation}
        \Omega^{(k)}_\beta := \left\{ b \in \Omega_0 \, : \quad 
          \overline{\varepsilon^{(k)}_n(\theta_{\mathbbm{R}}b) } = \beta_n \quad \mbox{for all $ n \in \mathbb{N}_0 $} \right\} 
      \end{equation}
      contains an almost-sure event which does not depend on the chosen wave number $ k \in \mathbbm{R} $. Therefore the super-event 
      is itself almost sure, $ \mathbbm{P}(\Omega^{(k)}_\beta) = 1 $ for all $  k \in \mathbbm{R} $. 
  \end{nummer}
\end{theorem}
\begin{remarks}
  \begin{nummer} 
    \item
      As a consequence of Theorem~\ref{thm:nreb}, 
      the pure-point spectrum
      and the absolutely continuous spectrum of $ H(b) $
      are also closed sets, $ \sigma_{\rm pp}(H(b)) = \bigcup_{|\beta_n| = 0 } \, \beta_n $ and 
      $ \sigma_{\rm ac}(H(b)) = \bigcup_{|\beta_n| > 0 } \, \beta_n $, 
      which do not depend on $ b \in \Omega_\beta $ (cf.\ Proposition~\ref{prop:spectral}).
    \item 
      The second part of Theorem~\ref{thm:nreb} deals with the distribution of the random variables $ b \mapsto 
      \varepsilon^{(k)}_n(b) $ for a fixed wave number $ k \in \mathbbm{R} $. 
      In view of the $ \mathbbm{R} $-ergodicity of $ \mathbbm{P} $, it is not surprising that
      the whole band $ \beta_n $ is explored by a single orbit
      $ \theta_{\mathbbm{R}}b := \{ \theta_{z_1} b \, : \, z_1 \in \mathbbm{R} \} \subset \Omega_0 $ 
      with $ \mathbbm{P} $-almost every ``initial'' $ b \in \Omega_0 $.
    \item
      Similarly to the energy bands, each \emph{asymptotic-velocity band} 
      \begin{equation}
        \overline{\Big] \inf_{k \in \mathbbm{R}} \frac{d \varepsilon^{(k)}_n(b)}{dk} ,
        \sup_{k \in \mathbbm{R}} \frac{d\varepsilon^{(k)}_n(b)}{dk} \Big[}, \qquad 
         n \in \mathbbm{N}_0, 
      \end{equation}
      is the same for $ \mathbbm{P} $-almost all $ b \in \Omega_0 $. 
      As a consequence, the spectrum of $ \overline{V}_{2,\infty} $ does not depend on  
      $ \mathbbm{P} $-almost all $ b \in \Omega_0 $.
      The proof of this statement is similar to that of Theorem~\ref{thm:nreb}.    
  \end{nummer}    
\end{remarks}
\begin{proof}[Proof of Theorem~\ref{thm:nreb}] 
  Shifting a realisation $ b \in \Omega_0 $ of a RUMF by $ z_1 \in \mathbbm{R} $ (cf.\ Remark~\ref{rem:ergod}) 
  implies the (covariance) relation 
  \begin{equation}
    \varepsilon^{(k)}_n(\theta_{z_1}b)=\varepsilon^{(k+ a(z_1))}_n(b)
  \end{equation}
  for the corresponding energy eigenvalues. As a consequence, for each $ n \in \mathbb{N}_0 $ the two 
  random variables $ b \mapsto \inf_{k \in \mathbb{R}} \varepsilon^{(k)}_n(b) $ and 
  $ b \mapsto \sup_{k \in \mathbb{R}}\varepsilon^{(k)}_n(b) $ are invariant under the action of 
  $ \{\theta_{z_1} \}_{z_1 \in \mathbbm{R}} $. 
  By the ergodicity there exists an event $ \Omega^{(n)} \subseteq \Omega_0 $  
  with $ \mathbbm{P}(\Omega^{(n)} ) = 1 $, 
  on which both random variables are constant \cite{CoFoSi82,Kal02}. 
  Since $ \bigcap_{n \in \mathbbm{N}_0} \Omega^{(n)}  \subseteq \Omega_\beta $ by virtue of (\ref{eq:band}) and 
  $ \mathbbm{P}\big(\bigcap_{n \in \mathbbm{N}_0} \Omega^{(n)} \big) = 1 $, this proves the first assertion.
  To prove the second one, we note that the continuity of $a(x_1) $ in $ x_1 \in \mathbbm{R} $  
  and (\ref{eq:mittel}) guarantee that 
  $ \mathbbm{P}(\widehat\Omega) = 1 $ for $ \widehat\Omega := \{ b \in \Omega \, : \,   a(\mathbbm{R}) = \mathbbm{R} \} $
  and hence
  \begin{equation}
    \varepsilon^{(\mathbbm{R})}_n(b) = 
    \varepsilon^{(k + a(\mathbbm{R}))}_n(b) =  
    \varepsilon^{(k)}_n(\theta_{\mathbbm{R}}b) 
  \end{equation}
  for all $ k \in \mathbbm{R} $ and all $ b\in \Omega_0 \cap \widehat\Omega  $. 
  This implies 
  $ \beta_n = \overline{\varepsilon^{(\mathbbm{R})}_n(b)} = \overline{\varepsilon^{(k)}_n(\theta_{\mathbbm{R}}b) } $
  for all $ n \in \mathbbm{N}_0 $ and all $ b $ in the almost-sure event $ \Omega_\beta \cap \widehat\Omega $.
\end{proof} 
\subsection{More on the energy bands in the sign-definite case}
Theorem~\ref{thm:loccont} guarantees that the energy eigenvalues $ \varepsilon^{(k)}_n(b) $ are continuous functionals 
of $ b \in \Omega_0 $ provided
the probability measure is concentrated on realisations with a definite sign.
This continuity has an important consequence. 
The energy bands turn out to be determined by any subset of $ \Omega_0 $ which is dense in the topological support 
of the probability measure. Such a subset may well be almost impossible or not even an event. 
\begin{theorem}[Subsets of the energy bands in the sign-definite case]\label{thm:detband}
Let $ (\Omega, \mathcal{B}(\Omega) , \mathbbm{P} )  $ be a RUMF for which there exists a constant $ b_- \in ]0,\infty[ $
such that the event
\begin{equation}
  \Omega_{b_-} := \big\{ b \in \Omega_0 \, : \, b(\mathbbm{R}) \subseteq  [b_-,\infty[ \big\}
\end{equation}
is almost sure, $ \mathbbm{P}(\Omega_{b_-}) = 1 $. 
Then 
\begin{enumerate}
        \item[(i)]
        $ \varepsilon_n^{(k)}(b) \in \beta_n  $~ for all 
        $ b \in \Omega_{b_-} \cap \mathrm{supp} \, \mathbbm{P} $;
        \item[(ii)]
         $ \overline{\varepsilon^{(k)}_n(\Delta)} = \beta_n $~ for all $ \Delta \subseteq \Omega_{b_-} \cap \mathrm{supp} \, \mathbbm{P} $ 
         with ~$ \overline{\Delta} = \mathrm{supp} \, \mathbbm{P} $
\end{enumerate}
for any band index $ n \in \mathbbm{N}_0 $ 
      and any wave number $ k \in \mathbbm{R} $.
\end{theorem}
\begin{remarks}
  \begin{nummer}
  \item
        We recall from Theorem~\ref{thm:loccont} that $ \beta_n \subseteq [(n+1/2) \, b_- , \infty [ $ for all $ n \in \mathbbm{N}_0 $ 
        in the situation of 
        Theorem~\ref{thm:detband}. 
    \item
        Theorem~\ref{thm:detband} and its proof below 
        is analogous to corresponding results for Schr\"o\-dinger operators with random scalar potentials
        \cite[Thms.~1 and 2 on p.~304f]{Kir89}.   
    \item\label{rem:constant}
        Theorem~\ref{thm:detband} 
        can be used to prove the almost-sure absence of flat energy bands of $ H(b) $. 
        Namely, to prove that $ \beta_n $ is not flat one has to track down two realisations $ b $, 
        $ b' \in \Omega_{b_-} \cap \mathrm{supp} \, \mathbbm{P} $ such that 
        $ \varepsilon_n^{(k)}(b) \neq \varepsilon_n^{(k)}(b') $ for some $ k \in \mathbbm{R} $. 
        This is the case, for example, if there are two constants $ b_0 > b_0' \geq b_- $ such that the constant 
        functions $ x_1 \mapsto b_0 $ and
        $ x_1 \mapsto b_0'  $ are both contained in $ \mathrm{supp} \, \mathbbm{P} $, see Corollary~\ref{cor:sqgauss} below.
  \end{nummer}
\end{remarks}
\begin{proof}[Proof of Theorem~\ref{thm:detband}] 
For fixed but arbitrary $ b \in \Omega_{b_-}\cap \mathrm{supp} \, \mathbbm{P} $ and $ \delta > 0 $ 
we have the strict positivity
$ \mathbbm{P}\big(\Delta_\delta(b) \cap \Omega_{b_-} \cap \Omega_\beta\big) = \mathbbm{P}(\Delta_\delta(b)) > 0 $
and therefore $ \Delta_\delta(b) \cap \Omega_{b_-} \cap \Omega_\beta \neq \emptyset $. 
By picking $ b_l \in \Delta_{1/l}(b) \cap \Omega_{b_-} \cap \Omega_\beta $ 
we can thus construct a sequence $ (b_l)_{l\in\mathbbm{N}} $ such that $ \lim_{l \to \infty} \mathrm{d}(b, b_l) = 0 $ and
hence $ \lim_{l \to \infty} \varepsilon^{(k)}_n(b_l) = \varepsilon^{(k)}_n(b) $ by Theorem~\ref{thm:loccont}.
Since 
$ \varepsilon^{(k)}_n(b) \in \overline{\bigcup_{l \in \mathbbm{N}}\{ \varepsilon^{(k)}_n(b_l)\} }\subseteq \beta_n $
by the definition~(\ref{eq:nreb}), this implies the first assertion. 
To prove the second one, we let $ z_1 \in \mathbbm{R} $ and 
$ b \in  \Omega_{b_-} \cap \Omega^{(k)}_\beta \cap \, \mathrm{supp} \,\mathbbm{P} $. 
Since all three events of the intersection are invariant under $ \theta_{z_1} $, we have $ \theta_{z_1} b \in  
\Omega_{b_-} \cap \Omega^{(k)}_\beta \cap \, \mathrm{supp} \,\mathbbm{P} $.
By the assumed denseness of  $ \Delta $ in $ \mathrm{supp} \, \mathbbm{P} $, 
there exists a sequence $(b_l)_{l\in\mathbbm{N}} $ 
with $ b_l \in \Delta $ such that  $ \lim_{l \to \infty} \mathrm{d}(\theta_{z_1}b, b_l) = 0 $ and
hence $ \lim_{l \to \infty} \varepsilon^{(k)}_n(b_l) = \varepsilon^{(k)}_n(\theta_{z_1}b) $
 by Theorem~\ref{thm:loccont}. 
Similarly as before, this implies $ \varepsilon^{(k)}_n(\theta_{z_1}b) \in \overline{\varepsilon^{(k)}_n(\Delta)} $. 
Since $ z_1 \in \mathbbm{R} $ was arbitrary and $ b \in \Omega^{(k)}_\beta $, Theorem~\ref{thm:nreb} gives 
$ \beta_n = \overline{\varepsilon^{(k)}_n(\theta_\mathbbm{R} b)} \subseteq \overline{\varepsilon^{(k)}_n(\Delta)} $.
This completes the proof, because $ \overline{\varepsilon_n^{(k)}(\Delta)} \subseteq \beta_n $ by assertion~(i).
\end{proof}
\subsection{On the absence of flat energy bands in the non-sign-definite case}
The following theorem provides a sufficient condition for the entire
spectrum of $ H(b) $ to be absolutely continuous and 
given by the positive half-line for  all $ b \in \Omega_\beta $.
According to Section~\ref{sec:UMF} 
the transport along the $ x_2 $-direction is then almost surely 
ballistic.
In fact, the condition 
guarantees the occurrence  of 
realisations $ b $ 
with arbitrarily small absolute values on spatial average  
over arbitrarily long line segments (cf.\ (\ref{eq:support}) and (\ref{eq:assumpconv})).
Not surprisingly, such realisations, which are rare because of our assumption $ \int_\Omega \mathbbm{P}(db) \, b(0) \neq 0 $, come with nearly free motion.
\begin{theorem}[Almost-sure absence of flat energy bands]\label{thm:crit}
  Let $ (\Omega, \mathcal{B}(\Omega) , \mathbbm{P} )  $ be a RUMF with the null-function of $ \mathrm{L}^1_{\mathrm{loc}}(\mathbbm{R}) $ lying
  in the topological support of its probability measure, $ 0 \in \mathrm{supp} \, \mathbbm{P} $. 
  Then 
  \begin{equation}\label{eq:abscont}
    \sigma(H(b) ) = \sigma_{\rm ac}(H(b) )   = [0, \infty[ 
  \end{equation}
  for all $ b \in \Omega_\beta $. 
\end{theorem}
\begin{remark}
  The almost-sure absolute continuity of the entire spectrum of $ H(b) $ implies that of its integrated density
  of states. This means that the \emph{density of states} exists as a non-negative function 
  in $ \mathrm{L}^1_{\rm loc}(\mathbbm{R} ) $
  (cf.\ \cite[Sec. 1.2]{LeMuWa03}). 
  For more general random vector potentials the integrated density of states is known to be only 
        H\"older continuous in certain energy regimes \cite{HiKl02}.  
\end{remark}
\begin{proof}[Proof of Theorem \ref{thm:crit}]
  To start the proof of the first equality in (\ref{eq:abscont}) by contradiction, 
  we note that zero cannot be an eigenvalue of the effective Hamiltonian $ H^{(0)}(b) $ (and hence $ \beta_0 \neq \{0\} $) 
  for all $ b \in \Omega_0 $, because 
  $ a^2 $ is strictly positive on some non-empty open set in $ \mathbbm{R} $ for all $ b \in \Omega_0 $.
  Suppose now that there exists an energy $ \varepsilon > 0 $ such that $  \beta_m = \{ \varepsilon \} $ 
  for some $ m \in \mathbbm{N}_0 $.  By (\ref{eq:assumpconv}) the assumption $ 0 \in \mathrm{supp} \, \mathbbm{P} $ 
  implies the existence of a sequence 
  $ (\Omega_l)_{l \in \mathbbm{N}}   $ 
  of non-empty events $ \Omega_l \subset \Omega_\beta $ such that 
  \begin{equation}
    \sup_{x_1 \in [-l, l ]} \, | a(x_1) | \leq \int_{-l}^l  d x_1 \, | b(x_1) | < l^{-1} 
  \end{equation}
  for all $ b \in \Omega_l $. By picking a $ b_l \in  \Omega_l \neq \emptyset $ 
  for each $ l \in \mathbbm{N} $ we can thus 
  construct a sequence $ \big(b_l\big)_{l \in \mathbbm{N}} $  such that 
  $  \lim_{l \to \infty} \big\| 2 H^{(0)}(b_l) \varphi - P_1^2 \, \varphi \big\| = 0 $
  for all $ \varphi \in \mathcal{C}_0^\infty(\mathbbm{R}) $. 
  According to \cite[Thm.~VIII.25]{ReSi1} the sequence of operators
  $ \big(H^{(0)}(b_l)\big)_{l \in \mathbbm{N}} $ hence converges to the free Hamiltonian $ P_1^2/2 $ on $ {\rm L}^2(\mathbbm{R} ) $ in the strong resolvent sense. 
  Using \cite[Thm.~VIII.24]{ReSi1} and \cite[Prob.~167 on p.~385]{ReSi4} this delivers the estimate
  \begin{equation}\label{eq:contr}
    \mathrm{tr} \;\chi_{[0, \, \varepsilon [} \left(  P_1^2/2 \right) \leq   \limsup_{l \to \infty } \; 
    \mathrm{tr} \;\chi_{[0, \, \varepsilon [}\big(H^{(0)}(b_l) \big)
      = m.
  \end{equation}
  Here the equality stems from the fact that 
  the number of eigenvalues of $ H^{(0)}(b) $ below $ \varepsilon $ equals $ m $ for all $ b \in \Omega_\beta $, 
  since $  \beta_m = \{ \varepsilon \} $ 
  by assumption.
  Inequality (\ref{eq:contr}) now contradicts the fact that the spectral projection 
  $ \chi_{[0, \varepsilon [} \left(  P_1^2 \right) $ is not a trace-class operator  for any $ \varepsilon > 0 $. 
  To prove the second equality in (\ref{eq:abscont}), we note that the inequality in (\ref{eq:contr}) 
  also implies that the number of eigenvalues of $ H^{(0)}(b_l) $
  below a fixed energy $ \varepsilon > 0 $ exceeds every given 
  number for $ l $ large enough. Hence $ \varepsilon \in \beta_n $ for all $ n \in \mathbbm{N}_0 $. 
  Since $ \varepsilon $ may be chosen arbitrarily small and $ \beta_n $ is closed, we thus have
  $ 0  \in \beta_n $ for all $ n \in \mathbbm{N}_0 $. This implies the assertion, because $ H(b) $ is 
  unbounded from above for all $ b \in \Omega_0 $.
\end{proof}

\subsection{Examples}\label{subsec:ex}
In this final subsection we are going to present three examples of a RUMF to which the general theory applies.
Our first example of a RUMF will be a Gaussian one in the sense of 
\begin{definition}[Gaussian RUMF]\label{def:gauss}
  A \emph{Gaussian random unidirectionally constant magnetic field} 
  is a RUMF $ ( \Omega, \mathcal{B}(\Omega), \mathbbm{P}) $ 
  with $ \widetilde{\mathbbm{P}}(\eta) $ 
  having the form
  \begin{equation}
    \exp\left\{ - i \mu \int_{\mathbbm{R}} dx_1 \, \eta(x_1) 
      - \frac{1}{2} \int_{\mathbbm{R} \times \mathbbm{R} } \mkern-10mu 
      dx_1 dy_1 \, \eta(x_1) \,
      c(x_1 - y_1) \, \eta(y_1) \right\}.
  \end{equation}
  Here $ \mu \in \mathbbm{R}\backslash\{0\} $ is a constant and  
  $  c: \mathbbm{R} \to \mathbbm{R}, \;  x_1 \mapsto c(x_1) = \int_\mathbbm{R} \widetilde c(d q)\,  e^{i q x_1} $ 
  is the Fourier transform of a positive and symmetric Borel measure $ \widetilde c $ on $ \mathbbm{R} $ with 
  $  0 < \widetilde c(\mathbbm{R}) < \infty $
  and no pure-point part in its Lebesgue decomposition.
\end{definition}
\begin{remark}\label{rem:gauss}
  It follows that $ \mu = \int_\Omega \mathbbm{P}(db) \, b(x_1) $ and 
  $ c(x_1-y_1) = \int_\Omega \mathbbm{P}(db) \, b(x_1) b(y_1) - {\mu}^2 $ for Lebesgue almost all $ x_1 , y_1 \in \mathbb{R} $,
  so that $ \mu $ is the mean-value and $ c $ the covariance 
  function of the Gaussian $ \mathbbm{P} $. 
  According to the Bochner-Khinchin theorem \cite{ReSi2,CoFoSi82} the Fourier representability of a (continuous) 
  covariance function required 
  in Definition~\ref{def:gauss} is no loss of generality.  
  According to the Fomin-Grenander-Maruyama theorem \cite{CoFoSi82,CrLe67} the measure $ \widetilde c $, known as
  the \emph{spectral measure} of $ \mathbbm{P} $, has no pure-point part in its Lebesgue decomposition, that is, 
  $ \widetilde c(\{q\}) = 0 $ for all $ q \in \mathbbm{R} $, if and only if $ \mathbbm{P} $ is $ \mathbbm{R} $-ergodic.
  By the Wiener theorem \cite{CoFoSi82,CyFr87} this is also equivalent to 
  $ \lim_{\ell \to \infty} \ell^{-1} \int_0^\ell d x_1 \, \big(c(x_1)\big)^2 = 0 $.
\end{remark}
An immediate consequence of Proposition~\ref{prop:gauss} in Appendix~\ref{app} below is
\begin{corollary}
  Theorem~\ref{thm:crit} applies to a Gaussian RUMF. 
\end{corollary}

Our second example is a RUMF with realisations $ b = b_- + \hat b^2 $ given by
the sum of a strictly positive constant $  b_- > 0 $ and the square of realisations $ \hat b $
of a Gaussian RUMF, so that Theorem~\ref{thm:detband} (and Remark~\ref{rem:constant}) is applicable.

\begin{definition}[Squared-Gaussian RUMF]
  A  \emph{squared-Gaussian random unidirectionally constant magnetic field with infimum $ b_- \in ]0,\infty[ $}
  is a RUMF $ ( \Omega, \mathcal{B}(\Omega), \mathbbm{P}) $ whose probability measure $ \mathbbm{P} $ 
  is defined in terms of a Gaussian RUMF
  $ ( \Omega, \mathcal{B}(\Omega), \mathbbm{P}_{\mu,c}) $ with mean-value $ \mu $ and covariance function $ c $ 
  by setting $ \mathbbm{P}(\Delta) := \mathbbm{P}_{\mu,c}\{ \hat b \in \Omega \, : \, 
  b_- + \hat b^2 \in \Delta \} $ for all $ \Delta \in \mathcal{B}(\Omega) $. 
\end{definition}
\begin{corollary}\label{cor:sqgauss}
  Let $ ( \Omega, \mathcal{B}(\Omega), \mathbbm{P}) $ be a squared-Gaussian RUMF with infimum $ b_- > 0 $.
  Then 
  \begin{equation}
    \beta_n = \big[  (n+1/2)\,  b_-, \infty[ 
 \end{equation}
 for all $ n \in \mathbbm{N}_0 $. Consequently, the entire spectrum of $ H(b) $ is absolutely continuous for $ \mathbbm{P} $-almost
 all $ b \in \Omega $.
\end{corollary}
\begin{proof}
  With the help of Proposition~\ref{prop:gauss} it can be shown that 
  the constant realisation $ x_1 \mapsto b_- + b_0^2 $ is contained in 
  $ \Omega_{b_-} \cap {\mathrm{supp}} \, \mathbbm{P} $ for every $ b_0 \in \mathbbm{R} $. 
  Theorem~\ref{thm:detband}(i) 
  thus implies $  (n+1/2) (b_- + b_0^2) \in \beta_n $ for all $ n \in \mathbbm{N}_0 $ (cf.\ Example~\ref{ex:const}).
\end{proof}

Our last example of a RUMF is a Poissonian one in the sense of

\begin{definition}[Poissonian RUMF] 
  A \emph{Poissonian random unidirectionally constant magnetic field} is is a RUMF 
  $ ( \Omega, \mathcal{B}(\Omega), \mathbbm{P}) $ 
  with $ \widetilde{\mathbbm{P}}(\eta) $ 
  having the form
  \begin{equation}
    \exp\left\{ - \varrho \int_\mathbbm{R} dx_1 \left( 1 - \exp\left\{-i \int_\mathbbm{R} dy_1 \eta(y_1) u(x_1 -y_1)\right\}
        \right)\right\}.
  \end{equation}
  Here  $ \varrho \in] 0,\infty[ $ is a constant and $ u: \mathbbm{R} \to \mathbbm{R} $ is a function 
  in $ {\rm L}^1(\mathbbm{R}) $ 
  satisfying $  \int_{\mathbbm{R}} \! d y_1 \, u(y_1) $\hspace{0pt}$ \neq 0 $. 
\end{definition}
\begin{remark}\label{rem:poisson}
  It follows that $ \mathbbm{P} $ is $ \mathbbm{R} $-ergodic and that
  $ 0 \neq  \varrho \int_{\mathbbm{R}} \! d y_1 \, u(y_1) = \int_\Omega \mathbbm{P}(db) b(0) $\hspace{0pt}$\leq 
  \int_\Omega \mathbbm{P}(db) |b(0)|\leq  \varrho \int_{\mathbbm{R}} \! d y_1 \, |u(y_1)| < \infty $. 
  Moreover, for every Poissonian RUMF there exists a Poissonian (random) measure $ \nu_\varrho : \Omega \times \mathcal{B}(\mathbbm{R}) \to  [0, \infty] $, $ (b, \Lambda{}) \mapsto \nu_\varrho(b,\Lambda{}) $ with intensity parameter $ \varrho $ such that 
  $ \mathbbm{P} $-almost every $ b \in \Omega $ 
  can be represented as 
  \begin{equation}
    b(x_1) = \int_\mathbbm{R} \nu_\varrho(b,dy_1)\, u(x_1-y_1) 
  \end{equation} 
  for Lebesgue-almost all 
  $ x_1 \in \mathbbm{R} $.  
  We recall that $ \nu_\varrho $
  is a random Borel measure on $ \mathbbm{R} $ which is almost surely only pure-point and positive-integer valued. 
  The random variable $ \nu_\varrho(\Lambda{}) : \Omega \to  [0, \infty] $, $ b \mapsto \nu_\varrho(b,\Lambda{}) $
  associated with $ \Lambda{} \in \mathcal{B}(\mathbbm{R}) $ is distributed according to Poisson's law 
  \begin{equation}\label{eq:Poisson}
     \mathbbm{P}\left(\big\{ b \in \Omega \, : \, \nu_\varrho(b,\Lambda{}) = m \big\}\right)
     = \frac{ \left( \varrho | \Lambda{} | \right)^m}{m!} \exp\left(-\varrho |\Lambda{} |\right), \qquad m \in \mathbbm{N}_0,
  \end{equation}
  so that $ \varrho $ may be interpreted as the mean spatial concentration of immobile magnetic impurities.
  Each single one is located ``completely at random'' on the real line where it creates a local magnetic field given by $ u $. 
\end{remark}

\begin{corollary}\label{Poi}
  Theorem~\ref{thm:crit} applies to a Poissonian RUMF.
\end{corollary}
\begin{proof}
  The triangle inequality, the Fubini-Tonelli theorem and the monotonicity
  $ \int_{-\ell}^\ell \! d x_1  $\hspace{0pt}$| u(x_1 - y_1 ) | \leq \int_{\mathbbm{R}}  d x_1 \, | u(x_1 ) | =:  \| u \|_{1} $, valid for all real $ \ell > 0 $, yield 
  \begin{align}\label{eq:integralab}
    \int_{-\ell}^\ell  \!\! d x_1 \, | b(x_1) | & 
    \leq \int_\mathbbm{R} \nu_\varrho(b,dy_1) \, \int_{-\ell}^\ell  \!\! d x_1 \, | u(x_1 - y_1 ) | \notag\\
    & \leq \nu_\varrho\left(b,[-r,r]\right) \| u \|_{1} +  u_{\ell, r}(b)
  \end{align}
  for arbitrarily picked $ r > 0 $. 
  Here we have introduced the two-parameter family of non-negative random variables $  u_{\ell, r} $ given by 
  $ u_{\ell, r}(b) := \int_{\mathbbm{R}\backslash [-r,r]}   \nu_\varrho(b,dy_1)  \int_{-\ell}^\ell   d x_1 \, | u(x_1 - y_1 ) | $.
  The Poissonian nature of $ \nu_\varrho $ implies that the two random variables 
  $ \nu_\varrho([-r,r]) $ and $  u_{\ell, r} $ are independent for all $ \ell $, $ r > 0 $.
  Inequality~(\ref{eq:integralab}) therefore gives the following lower estimate on the probability for 
  the $ \delta $-smallness of its left-hand side:
  \begin{multline}\label{eq:proppoiss}
     \mathbbm{P}\Big(\Big\{ b \in \Omega \, : \, \int_{-\ell}^\ell  \!\! d x_1 \, | b(x_1) | < \delta \Big\} \Big)  \\
    \geq 
     \mathbbm{P}\left(\left\{ b \in \Omega \, : \, \nu_\varrho\left(b,[-r,r]\right)  \| u \|_{1} < \frac{\delta}{2}\right\} \right) 
    \; \mathbbm{P}\left(
\left\{ b \in \Omega \, : \, 
       u_{\ell, r}(b) < \frac{\delta}{2}\right\}\right). 
  \end{multline}
  The first probability on the right-hand side is strictly positive for all $ r > 0 $ by (\ref{eq:Poisson}) with $ m = 0 $.
  We estimate the second probability from below by bounding the probability of the complementary event from above as follows 
  \begin{align}
     \mathbbm{P}\left(\left\{ b \in \Omega \, : \, 
       u_{\ell, r}(b) \geq \frac{\delta}{2}\right\}\right) 
   & \leq \frac{2}{\delta} \; \int_\Omega\! \mathbbm{P}(db)\, u_{\ell, r}(b) \notag \\
   &   = \frac{2\varrho}{\delta} \int_{\mathbbm{R}\backslash [-r,r]} \mkern-20mu d y_1  \int_{-\ell}^\ell   d x_1 \, | u(x_1 - y_1 ) |.
      \label{eq:cheb}
  \end{align}
  Here we have used the Chebyshev-Markov inequality, the Fubini-Tonelli theorem and the identity 
  $  \int_\Omega \mathbbm{P}(db) \nu_\varrho(b,\Lambda{})= \varrho | \Lambda{} | $ for the mean number of Poissonian points in 
  $ \Lambda{} \in \mathcal{B}(\mathbbm{R}) $. 
  The right-hand side of (\ref{eq:cheb})
  becomes arbitrarily small with $ r $ large enough for any pair $ \delta $, $ \ell > 0 $ because $ u \in \mathrm{L}^1(\mathbbm{R})$. 
  Therefore the probability on the left-hand side of (\ref{eq:proppoiss}) is strictly positive for any $ \delta $, $ \ell > 0 $. 
  Hence the constant realisation $ b = 0 $ belongs to $ \mathrm{supp} \, \mathbbm{P} $ (cf.\ (\ref{eq:support}) and 
  (\ref{eq:assumpconv})). 
\end{proof}

\begin{remark}

In this paper we have only considered random UMF's which are $ \mathbbm{R}$-ergodic (by definition).
 But the results can 
easily be extended to certain random UMF's, which are not $\mathbbm{R}$-ergodic but only $\mathbbm{Z}$-ergodic. 
For example, if $ \widetilde{\mathbbm{P}}(\eta) $ has the form
\begin{equation}
  \prod_{j \in \mathbbm{Z}} \, \int_{\mathbbm{R}} \lambda(dg) \, \exp\left\{ -i g \int_{\mathbbm{R}} d x_1 \, \eta(x_1) \, 
    u(x_1 - j) \right\}  
\end{equation}
where $ \lambda $ is a probability measure on $ (\mathbbm{R}, \mathcal{B}(\mathbbm{R})) $ with $ 0 \in {\mathrm{supp}} \, \lambda $
and $ 0 < | \int_\mathbbm{R} \lambda(dg) g | $\hspace{0pt} $< \infty$, and $ u : \mathbbm{R} \to \mathbbm{R} $ is a function 
in $ {\mathrm{L}}^1(\mathbbm{R}) $ satisfying $\int_{\mathbbm{R}} dy_1 \, u(y_1) \neq 0$.  
Then $ \mathbbm{P} $-almost every realisation $ b $ can be represented as 
$ b(x_1) = \sum_{j \in \mathbbm{Z}} g_j(b) u(x_1 - j) $ for Lebesgue-almost all $ x_1 \in \mathbbm{R} $ in terms
of $ u $ and a two-sided sequence $ ( g_j )_{j \in \mathbbm{Z}} $ 
of independent random variables with common distribution $ \lambda $ and can easily be shown to be a UMF in the sense of
Definition~\ref{ass}. 
The assertions of Theorem~\ref{thm:nreb} and Theorem~\ref{thm:crit}
remain true for $ \mathbbm{P} $-almost all realisations $ b $ of this $\mathbbm{Z}$-ergodic random UMF. 
The proof of the latter statement is in close analogy to that of Corollary \ref{Poi}.
\end{remark}

\appendix
\section{On the topological support of certain Gaussian path measures}\label{app}
For any Gaussian RUMF $ (\Omega, \mathcal{B}(\Omega), \mathbbm{P}) $ in the sense of Definition~\ref{def:gauss} the event
\begin{equation}
  \Omega_2 := \left\{ b \in \Omega \, : \, b \in \mathrm{L}^2_{\mathrm{loc}}(\mathbbm{R}) \right\} 
  =  \left\{b \in \mathrm{L}^2_{\mathrm{loc}}(\mathbbm{R})\, : \, b(\mathbbm{R}) \subseteq \mathbbm{R} \right\} 
\end{equation}
is almost-sure, $ \mathbbm{P}(\Omega_2) = 1 $, 
because the Fubini-Tonelli theorem and the
$ \mathbbm{R} $-homogeneity of $ \mathbbm{P} $ gives 
$ \int_\Omega \mathbbm{P}(db) \int_{-\ell}^\ell dx_1 | b(x_1)|^2 = 2 \ell ( \mu^2 + c(0) ) < \infty $ for all real $ \ell > 0 $.  
It is therefore natural to consider the $ \mathrm{L}^2_{\mathrm{loc}} $-topological support
\begin{equation}
  \mathrm{supp}_2 \, \mathbbm{P} := \Big\{ b \in \Omega_2 \, : \, 
    \mathbbm{P}\left(\left\{ b' \in \Omega_2  \, : \, \mathrm{d}_2(b,b') < \delta \right\}\right)  > 0 \quad
    \mbox{for all $ \delta > 0 $} \Big\}
\end{equation}
associated with the metric on $ \Omega_2 $ defined by $ \mathrm{d}_2(b,b') : = 
\sum_{j \in \mathbbm{Z}} 2^{-|j|}  \, \min\big\{ 1 , \big(\int_j^{j+1} dx_1 $\hspace{0pt}$| b(x_1) -  b'(x_1) |^2 \big)^{1/2} \big\} $. 
Since $ \mathrm{d}(b,b') \leq \mathrm{d}_2(b,b') $ for all $ b$, $ b' \in \mathrm{L}^2_{\mathrm{loc}}(\mathbbm{R}) $, this
      $ \mathrm{L}^2_{\mathrm{loc}} $-topological support of $ \mathbbm{P} $ is contained in its
      ($ \mathrm{L}^1_{\mathrm{loc}} $-)topological support as given by (\ref{eq:support}). \\

Now we are able to recall a known fact (cf.\ \cite[p.\ 451]{Kot85}), 
which is actually valid for slightly more general Gaussian processes than Gaussian RUMF's. Its detailed proof is included here  
for the reader's (and authors') convenience.

\begin{proposition}[Topological support of a Gaussian RUMF]\label{prop:gauss}
For any Gaussian RUMF $ (\Omega, \mathcal{B}(\Omega), \mathbbm{P}) $ one has 
$ \Omega_2 = \mathrm{supp}_2 \, \mathbbm{P}  \; ( \subseteq \mathrm{supp} \, \mathbbm{P} ) $.
\end{proposition}
To prepare a proof we first recall the Karhunen-Lo\`eve expansion \cite{Adl90} of Gaussian processes. 
It relies on the fact that for each fixed $ \ell \in ]0, \infty[ $ the covariance 
function defines a non-negative and compact integral operator $ C $ on the Hilbert space $ {\rm L}^2([-\ell, \ell]) $ 
through the kernel  
$ [-\ell, \ell\, ]^2 \, \ni (x_1, y_1) \mapsto c(x_1-y_1) $. Mercer's theorem \cite{BeShUs90} therefore yields the existence of 
a basis of continuous real-valued eigenfunctions $( \phi_j)_{j \in \mathbbm{N}_0} $ which is orthonormal,
$ \langle \phi_j, \phi_l \rangle_{\ell} = \delta_{jl} $ for all $ j$, $l \in \mathbbm{N}_0$, 
with respect to the usual scalar product on $ {\rm L}^2([-\ell, \ell]) $ such that 
\begin{equation}\label{eq:mercer}
  c(x_1-y_1) = \sum_{j = 0}^\infty c_j \, \phi_j(x_1) \phi_j(y_1).
\end{equation}
Here $  c_0 \geq c_1 \geq  \dots \geq 0 $ are the corresponding 
non-negative (possibly coinciding) eigenvalues and the convergence of the series is absolute and uniform on 
the square $ [-\ell, \ell \, ]^2 \subset \mathbbm{R}^2$.
One even has $ c_j > 0 $ for all $ j \in \mathbbm{N}_0 $, if the spectral measure $ \widetilde c $ has a continuous part 
in its Lebesgue decomposition (as is the case for a Gaussian RUMF because of ergodicity). 
This follows from the strict positivity of the quadratic 
form associated with $ C $. Namely, the assumption $ \langle \varphi , C \varphi \rangle_{\ell}  
= \int_\mathbbm{R} \widetilde c(dq) | \widetilde \varphi(q) |^2 = 0 $ implies 
$ \widetilde \varphi(q) := \int_{-\ell}^\ell dx_1 e^{- i q x_1 } \varphi(x_1) = 0 $ for all $ \varphi \in  {\rm L}^2([-\ell, \ell]) $
and all $ q \in {\rm supp} \, \widetilde c := \{q \in \mathbbm{R} \, : \,  
  \widetilde c(]q - \kappa , q + \kappa[) > 0 \;\; \mbox{\rm for all $ \kappa > 0 $} \} $.
Since $ | {\rm supp} \, \widetilde c | > 0 $ by the assumed existence of a continuous part of $ \widetilde c $, 
the analyticity of the complex-valued 
function $ \mathbbm{R} \ni q \mapsto \widetilde \varphi(q) $ 
implies $ \widetilde \varphi(q) = 0 $ 
even for all $ q \in \mathbbm{R} $ and hence $ \varphi = 0 $. 
 
Using (\ref{eq:mercer}) we can define a sequence $ (\gamma_j)_{j \in \mathbbm{N}_0} $ of (jointly) Gaussian random variables by
\begin{equation}
  \gamma_j(b) := \int_{-\ell}^\ell \!\! d x_1 \;  \phi_j(x_1)  \left( b(x_1)- \mu \right), \quad b \in \Omega.
\end{equation}
They have zero mean-values, have strictly positive variances and are pairwise uncorrelated, 
$ \int_\Omega \mathbbm{P}(db)  \gamma_j(b) = 0 $ and $ \int_\Omega \mathbbm{P}(db) \,\gamma_j(b) \gamma_l(b) = c_j \delta_{jl} $
for all $ j $, $ l  \in \mathbbm{N}_0 $. By their Gaussian nature, they are thus independent \cite{Kal02}.

\begin{proof}[Proof of Proposition~\ref{prop:gauss}] 
  Inequalities analogous to (\ref{eq:assumpconv}) show that $ \hat b \in \mathrm{supp}_2 \, \mathbbm{P} $ if and only if
    \begin{equation}\label{eq:gauss}
      \mathbbm{P}\left(\big\{ b \in \Omega_2 \, : \,  \| b - \hat b \|_{2,\ell} < \delta \big\}\right) > 0 
    \end{equation}
    for all $ \delta > 0 $ and all $ \ell > 0 $. Here we have introduced the abbreviation
    $ \| b  \|_{2,\ell}^2 := \int_{-\ell}^\ell dx_1 | b(x_1) |^2 $ for the squared norm 
    of $ b \in  {\rm L}^2([-\ell, \ell]) $.
  For a proof of (\ref{eq:gauss}) for arbitrary $ \hat b \in \Omega_2 $, 
  we may assume $ \mu= \int_\Omega \mathbbm{P}(db) \, b(0) = 0 $ by adding a suitable constant to $ \hat b $. 
 We $ {\rm L}^2([-\ell, \ell]) $-expand with respect to 
  the basis $ (\phi_j )_{j \in \mathbbm{N}_0 } $ 
  and employ the triangle inequality to obtain 
  \begin{align}
      & \big\|  b - \hat b \big\|_{2,\ell} 
      =    \Big( \sum_{j= 0}^\infty \big| \gamma_j(b) - \langle \phi_j , \hat b \rangle_\ell \big|^2 \Big)^{1/2} \\
      &\;  \leq   \Big( \sum_{j= 0}^{m-1}   \big| \gamma_j(b) -  \langle \phi_j , \hat b \rangle_\ell \big|^2 \Big)^{1/2}  
     + \Big( \sum_{j= m}^\infty  |\gamma_j(b)|^2 \Big)^{1/2} 
     +  \Big( \sum_{j= m}^\infty \big| \langle \phi_j , \hat b \rangle_\ell \big|^2 \Big)^{1/2} \notag
  \end{align}
  for any $ m \in \mathbbm{N} $. Now, given $\delta > 0$, the last term does not exceed $\delta/3$ for $ m $ large enough, because 
  of Parseval's identity  
  $ \sum_{j=0}^\infty | \langle \phi_j , \hat b \rangle_\ell |^2 = \| \hat b \|_{2, \ell}^2  < \infty $.
  By the independence of the $ (\gamma_j) $, for all $m$ large enough the probability in (\ref{eq:gauss}) is therefore bounded from below by the following product of two probabilities:
  \begin{equation}\label{eq:prodprob}
    \mathbbm{P}\Big(\Big\{ b \in \Omega_2 \, : \, 
     \sum_{j= 0}^{m-1}   \big| \gamma_j(b) -  \langle \phi_j , \hat b \rangle_\ell \big|^2 < \frac{\delta^2}{9}   \Big\}\Big) 
      \;  \mathbbm{P}\Big(\Big\{b \in \Omega_2 \, : \, \sum_{j= m}^\infty  |\gamma_j(b)|^2 < \frac{\delta^2}{9}   \Big\}\Big). 
  \end{equation}
  The second probability in (\ref{eq:prodprob}) becomes strictly positive for all $m$ large enough, because the Chebyshev-Markov inequality and the convergence $ \sum_{j=0}^\infty c_j = 2 \ell\,  c(0) =  2 \ell\,  \widetilde c(\mathbbm{R}) < \infty $ then yield
  \begin{equation} 
  \mathbbm{P}\Big(\Big\{ b \in \Omega_2 \, : \, \sum_{j= m}^\infty  |\gamma_j(b)|^2 \geq \frac{\delta^2}{9}  \Big\}\Big) 
  \leq \frac{9}{\delta^2} \sum_{j= m}^\infty c_j < 1.
  \end{equation}
  It remains to ensure the strict positivity of the first probability in (\ref{eq:prodprob}). 
  By the independence of the Gaussian random variables $ (\gamma_j) $ one has
  \begin{multline}
    \mathbbm{P}\Big(\Big\{ b \in \Omega_2 \, : \,   
    \sum_{j=0}^{m-1} \Big| \gamma_j(b) -   \langle \phi_j , \hat b \rangle_\ell \Big|^2 < \frac{\delta^2}{9} \Big\}\Big)  \\
    \geq  \prod_{j=0}^{m-1} \; \mathbbm{P}\left(\Big\{ b \in \Omega_2 \, : \, 
   \big| \gamma_j(b) -  \langle \phi_j , \hat b \rangle_\ell \big| < \frac{\delta}{3 \sqrt{m}}  \Big\}\right).
   \label{1dG}
  \end{multline} 
  Since $ c_j > 0 $ for all $ j \in \mathbbm{N}_0 $, each of the $ m $ probabilities 
  on the right-hand side of (\ref{1dG}) 
  is strictly positive, because
  a Gaussian probability measure on $ (\mathbbm{R}, \mathcal{B}(\mathbbm{R})) $ with strictly positive variance assigns a strictly positive value to any non-empty open interval. 
\end{proof}

\section*{Acknowledgements}
We are indebted to Karl Petersen (Chapel Hill, North Carolina), 
J\"urgen Potthoff (Mannheim, Germany), Michael R\"ockner (Bielefeld, Germany) and 
Ludwig Schweitzer (Braunschweig, Germany) for helpful advice and hints to the literature. 
This work was partially supported by the Deutsche Forschungsgemeinschaft (DFG) under grant nos. Le 330/12 and Wa 1699/1.
The former is a project within the DFG Priority Programme SPP 1033 ``Interagierende stochastische Systeme von hoher
Komplexit\"at''.


\end{document}